\documentclass[pra,letterpaper,aps,10pt,superscriptaddress,twocolumn,floatfix]{revtex4-1}
\usepackage{amsmath,graphicx,amssymb,braket,xcolor,subfigure,upgreek}
\usepackage[colorlinks, linkcolor=blue, citecolor=blue, urlcolor=blue, breaklinks=true]{hyperref}
\usepackage{empheq,verbatimbox}

\newcommand{\twotwomat}[4]{\begin{bmatrix}#1&#2\\#3&#4\end{bmatrix}}
\newcommand{\sgn}{\text{sgn}}

\begin{document}

\title{Transmissive optomechanical platforms with engineered spatial defects}

\author{Edoardo Tignone}
\author{Guido Pupillo}
\affiliation{ISIS (UMR 7006) and IPCMS (UMR 7504), Universit\'{e} de
Strasbourg and CNRS, Strasbourg, France}
\author{Claudiu Genes}
\affiliation{Institut f\"ur Theoretische Physik, Universit\"at
Innsbruck, Technikerstrasse 25, A-6020 Innsbruck, Austria}

\date{\today}

\begin{abstract}
We investigate the  optomechanical photon-phonon coupling of a single light mode propagating through an array of vibrating mechanical elements. As recently shown for the particular case of a periodic array of membranes embedded in a high-finesse optical cavity [A. Xuereb, C. Genes and A. Dantan, Phys. Rev. Lett., \textbf{109}, 223601, (2012)], the intracavity linear optomechanical coupling can be considerably enhanced over the single element value in the so-called \textit{transmissive regime}, where for motionless membranes the whole system is transparent to light. 
Here, we extend these investigations to quasi-periodic arrays in the presence of engineered spatial defects in the membrane positions.  In particular we show that the localization of light modes induced by the defect combined with the access of the transmissive regime window can lead to additional enhancement of the strength of both linear and quadratic optomechanical couplings.

\end{abstract}


\maketitle

\section{Introduction}

\begin{figure*}[t]
 \includegraphics[width=1.99\columnwidth]{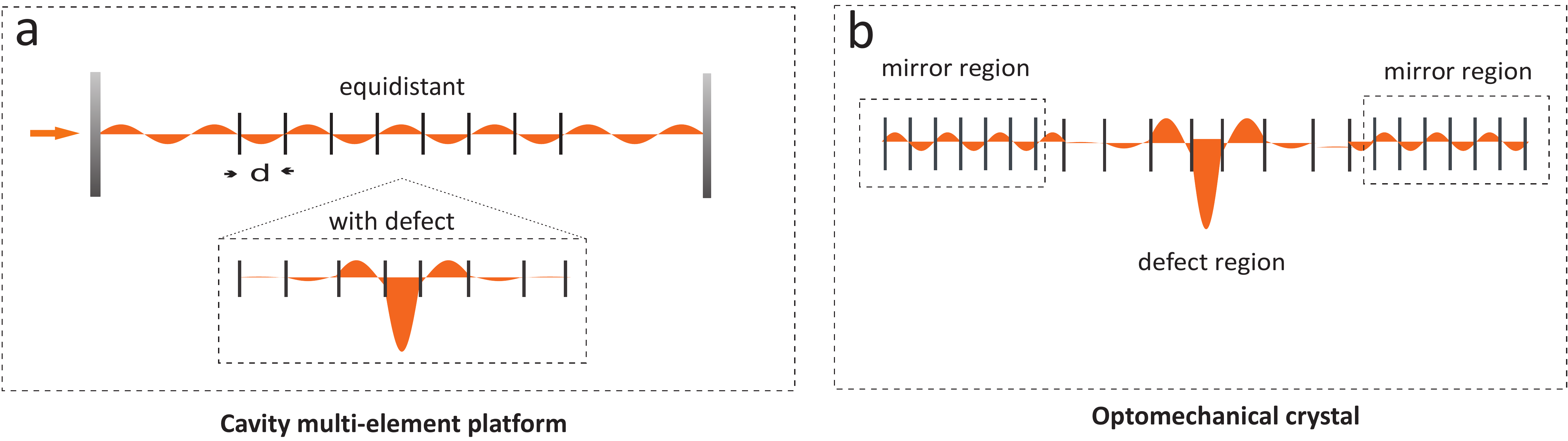}
 \caption{\emph{Optomechanical platforms} a) Cavity optomechanics with a transmissive equidistant membrane array
 shows localization of light within the middle region of the array when quadratic spacing defects are introduced. b) A simple 1D model of an OM crystal divided in 3 regions, the side ones are modelled as effective dispersive mirrors while the middle one accommodates the quadratic
 defect.}
 \label{fig:fig1}
 \end{figure*}

Cavity optomechanics (OM) both at the large mass scale (mirrors, membranes, levitated nano-particles, etc.)~\cite{asp13} and at the small mass scale (atoms, molecules, ions, etc.)~\cite{rit13} generally describes the classical and quantum
dynamics of systems of mobile scatterers manipulated via the
interaction with optical fields resonantly amplified by use of
end-mirrors (the so-called Fabry-P\'{e}rot cavity setup). Effects such as
cooling, heating, limit cycles or bistability occur owing to the
time delay between OM action and back-action, which
springs from the inherent timescale imposed by the cavity damping
process. Typical OM systems span over many orders of
magnitude in mass: at the microscopic
level cavity cooling of atoms has been proposed and experimentally tested more than a decade ago
~\cite{rit97,vul00,mau04}, while, at the other end of the spectrum, cavity
cooling of mirrors~\cite{gro09}, microtoroids~\cite{ver12}, sub-micron dielectric beads~\cite{kie13,ase13}
is a more recent endeavour. A long-sought goal is quantum
control of the OM interaction at the single
photon-phonon level (allowing, e.g., the
engineering of a coherent quantum interface between light and
motion)~\cite{akr10,rab11,nun11,nun12,qia12,lud12,kro13}. In the nonlinear regime, strong
OM interactions with a quadratic position dependence could allow for quantum
non-demolition detection of single phonon quantum
jumps at the macroscopic level, as for example in the so-called membrane in the middle setup~\cite{jay08,tho08}.

While most works have so far focused on OM platforms employing a single mechanical element, recent studies have started to explore multi-element approaches both theoretically~\cite{xue12,xue13,eis04,che14,bha08,har08,lud10,dob10,sta10,hei11,cha11,sta12,seo12,lud13} and experimentally~\cite{lin10,mah11,mah12,mas11,zha12}. In this case, it has been shown that large collective mechanical effects can occur when multiple scatterers are
addressed by a common light field. For example, recent experiments with systems of $N$ atoms in a cloud~\cite{bra12}, or trapped in optical lattices~\cite{cam11} have demonstrated enhanced linear couplings to light fields where the interaction strength scales with the atom number as $\sqrt N$, which is typical for center-of-mass addressing. Experiments on quasi-periodic dielectric media such as photonic crystals with engineered quadratic defects
have also shown a tremendous increase in photon-phonon linear coupling; this stems from the induced localization of fields within a very
small volume inside the crystal and the subsequent activation of mechanical collective modes defined by the defect and localized within the same small volume~\cite{eic09,cha09}. Alternatively, recent theoretical proposals have demonstrated that linear  couplings can be largely enhanced for OM cavities made of membrane arrays prepared in the {\it transmissive regime}. Here the whole system is essentially transparent for the ``frozen" configuration, while vibrations of the mechanical elements lead to strong phase shifts of the optical field. As opposed to the case of the OM crystals, a variety of collective mechanical modes with sinusoidal profiles are defined by the operational points in the transmission window. For $N$ membranes, there are $N-1$ such sinusoidal collective modes, each of them exhibiting a strong linear OM coupling that scales favorably with both membrane polarizability $\zeta$ and their number, as $\zeta^2 N^{3/2}$~\cite{xue12,xue13}.

In this paper we combine different approaches to investigate 1D (one-dimensional) OM systems - or {\it optomechanical
platforms} - obtained by specific designs of quasi-periodic multiple
scatterer media. In particular we focus on two platforms: (i) a
system made of a periodic array of membranes surrounded by an optical cavity (as introduced in Ref.~\cite{xue12,xue13}), where the array periodicity is modified by adding a ``defect" as a quadratic displacement of the membranes; in this case we look for enhancement in both linear and
quadratic couplings between light and the collective mechanical motion of the membranes with respect to the equidistant case; (ii) a simplified model for an OM crystal where we apply a 1D transfer matrix formalism to explore the possibility of exploiting the transmissive OM method. As a main result of our investigations, we show that platform (i) can exhibit increased OM linear and quadratic couplings (owing to the presence of the defect) with respect to the equidistant case treated in~\cite{xue12}. We then shift the discussions from mechanical to optical degrees of freedom and
remark that the similarity between the Helmoltz and
Schr\"odinger equations allows one to get more insight into the mechanism of light modes localization~\cite{rey05,qui04}.

The paper is organized as follows. In the next section we introduce the transfer matrix formalism and the two
platforms that we investigate. Section~\ref{sec:secIII} is devoted to the
study of the cavity-embedded membrane array where an analytically
solvable regime is identified corresponding to small defects. Outside this regime, general numerical
investigations are carried out for both linear and quadratic OM couplings. The discussion shifts to OM crystals in
Sec.~\ref{sec:secIV} where we analytically describe the system as an effective optical cavity with
dispersive mirrors that allow for the tuning of the resonances. In Sec.~\ref{sec:outlook} we offer a simple
interpretation of the physics of localization of light modes between
membranes by performing a mapping of the Helmoltz equation onto the
Schr\"odinger equation. Finally, Sec.~\ref{sec:conclusions}
concludes the paper.
\section{Model}
We consider non-absorptive optical elements such as membranes or
mirrors modeled as scatterers which in a 1D approach
are completely characterized by the real negative susceptibility
denoted by $\zeta$. The corresponding amplitude reflectivity is
$r=i\zeta/(1-i\zeta)$. Assuming an electric field of the form $E(x,t)=E(x)e^{i\omega t}$, with $x$ and $t$  the spatial and temporal coordinates, respectively, and $\omega$ the field frequency, the 1D wave equation describing the interaction of the field
with a single fixed beam splitter positioned at the origin ($x=0$) corresponds to the following Helmholtz equation
\begin{equation}
\label{eq:waveq}
\left[\partial_x^2 +\left(\frac{\omega}{c}\right)^2\epsilon_r(x)\right]E(x)=0,
\end{equation}
with $c$ the speed of light.
The relative permittivity can be decomposed as $\epsilon_r(x)=\epsilon_{r0}+\delta\epsilon_r(x)$, where $\epsilon_{r0}$ is the relative permittivity of the vacuum, while
\begin{equation}
\delta\epsilon_r(x)=\frac{2}{k}\zeta\delta(x),
\label{eq:eq2}
\end{equation}
with $k$ the wave vector of the light field.

\subsection{The transfer matrix approach}
The 1D problem of light propagation through an ensemble
of scatterers obeys the afore-mentioned Helmholtz equation. Here, we consider a discrete medium where each
optical element is infinitely thin and its position is labeled by an index $i$. The problem can be analyzed using
the transfer matrix formalism~\cite{deu95,xue09}, corresponding to a {\it beam-splitter}-type approach. The electric field at any point can be written as a vector where the two entries are the amplitudes of its left and right propagating components. We thus proceed by writing the
left and right traveling waves at the left and right of element $i$
as vectors, $v^{\pm}_i=(L^{\pm}_i, R^{\pm}_i)^\top$. These vectors are connected
by the following two matrices
\begin{equation}
\label{eq:M}
M_i = \twotwomat{1+i\zeta_i}{i\zeta_i}{-i\zeta_i}{1-i\zeta_i}\,,
\end{equation}
describing scattering at the mechanical element $i$ (such that $v^{-}_i= M_i
v^{+}_i$), and
\begin{equation}
\label{eq:F}
F_{i,i+1} = \twotwomat{e^{ikd_{i,i+1}}}{0}{0}{e^{-ikd_{i,i+1}}}\,,
\end{equation}
which describes propagation of a monochromatic beam with a wave number $k$
over a distance $d_{i,i+1}$ through free space (such that
$v^{+}_i= F_{i,i+1} v^{-}_{i+1}$).

\subsection{Two optomechanical platforms}
\label{sec:toa}
Let us consider the two distinct OM systems illustrated
in Fig.~\ref{fig:fig1}. In both cases we start by positioning $N$
membranes with polarizability $\zeta$ around the origin as
\begin{equation}
\label{eq:equipos} x^0_j=D
\left(-\frac{1}{2}+\frac{j-1}{N-1}\right),
\end{equation}
such that they are equidistant and separated by a distance
$d=D/(N-1)$. The ensemble of membranes thus constitutes a total optical discrete medium of
 length $D$. We then introduce a quadratic defect in the
spatial separation between neighboring membranes, by pushing them
progressively towards the origin while keeping the total length $D$
fixed. The position $x_j$ of element $j$ is thus
\begin{equation}
\label{eq:quadpos} x_j=x^0_j-\frac{\alpha}{d}
\left(\frac{D^2}{4}+{x^0_j}^2\right)\sgn{(x^0_j)},
\end{equation}
where $\alpha$ is smaller than $2/[N(N-1)]$ or $4/[(N-3)(N+1)]$ (for $N$
even or odd, respectively). \\On the first platform (see Fig.~\ref{fig:fig1}a) we position two mirrors
at $\pm L/2$, forming an optical cavity. The mirrors are placed far enough from the array ($L \gg D$) such that the finesse of the cavity can be very large.
On the second platform (see Fig.~\ref{fig:fig1}b), instead,  we place an array of $N_m$ membranes (with polarizability $\zeta_m$) on the left and right of membranes $1$ and $N$, respectively. These membranes are separated from the central array and from each other by a distance $d_m$. The two major differences
between these two platforms are that: (i) $L$ is a free parameter in platform
\textbf{a} allowing one to manipulate the free spectral range of the
optical cavity, e.g., to make it much smaller than the typical range in
which the optical response of the array in the middle varies
strongly; (ii) the dispersion relation (reflectivity function of the
wave vector $k$) is fixed for the side mirrors for platform \textbf{a} while
it is adjustable and controllable for platform \textbf{b}.

To describe the OM coupling regimes, we allow membranes
$1$ to $N$ to oscillate around their equilibrium positions and
quantify the changes induced by these oscillations on the resonances of the whole optical platforms. We identify these resonances by computing their wave vector $k$ as well as their first and second derivatives with respect to
small displacements
\begin{equation}\label{eq:eqCoup}
    g^{(1)}_j=c\frac{\delta k}{\delta x_j}x_0 \;\;\; \text{and} \;\;\; g^{(2)}_j=c \frac{\delta^2 k}{\delta x_j^2}x_0^2.
\end{equation}
The quantities $g^{(1)}_j$ and $g^{(2)}_j$ define the linear and quadratic couplings of the light field to the mechanical element $j$. Reference values are those computed for a typical single-element ``membrane-in-the-middle" setup:

\begin{equation}
\label{eq:std} g^{(1)}_0=\frac{2 c
k}{L}\frac{\zeta}{\sqrt{1+\zeta^2}}x_0 \;\;\; \text{and} \;\;\;
g^{(2)}_0 = \frac{2 c k^2}{L}\zeta x_0^2.
\end{equation}
In the following we define the optimal value $g^{(1)}_0=2 c k x_0/L\equiv g$, reached for unit reflectivity.

\section{Array of membranes}
\label{sec:secIII}

Replacing the single membrane with a multielement discrete optical medium (comprised of $N$ membranes each with polarizability $\zeta$, as described above) with tunable optical properties (e.g., such as reflectivity dependence on inter-element spacing) has been shown to lead to an improved scaling of the linear OM coupling, both with $\zeta$ and $N$ far above the single-membrane optimal coupling $g$~\cite{xue12}.
In the following, we first review these results  (Sec.~\ref{subsec:IIIA}) and then analyze the modification to the system properties obtained by adding a quadratic defect to the inter-element spacing (Sec.~\ref{subsec:IIIB}). As illustrated in Fig.~\ref{fig:fig2}, the optical response of the free-space membrane array, is already strongly modified from the equidistant case as the defect increases.

\subsection{Equidistant array}
\label{subsec:IIIA} 
\begin{figure}[t]
 \centering
 \includegraphics[width=0.80\columnwidth]{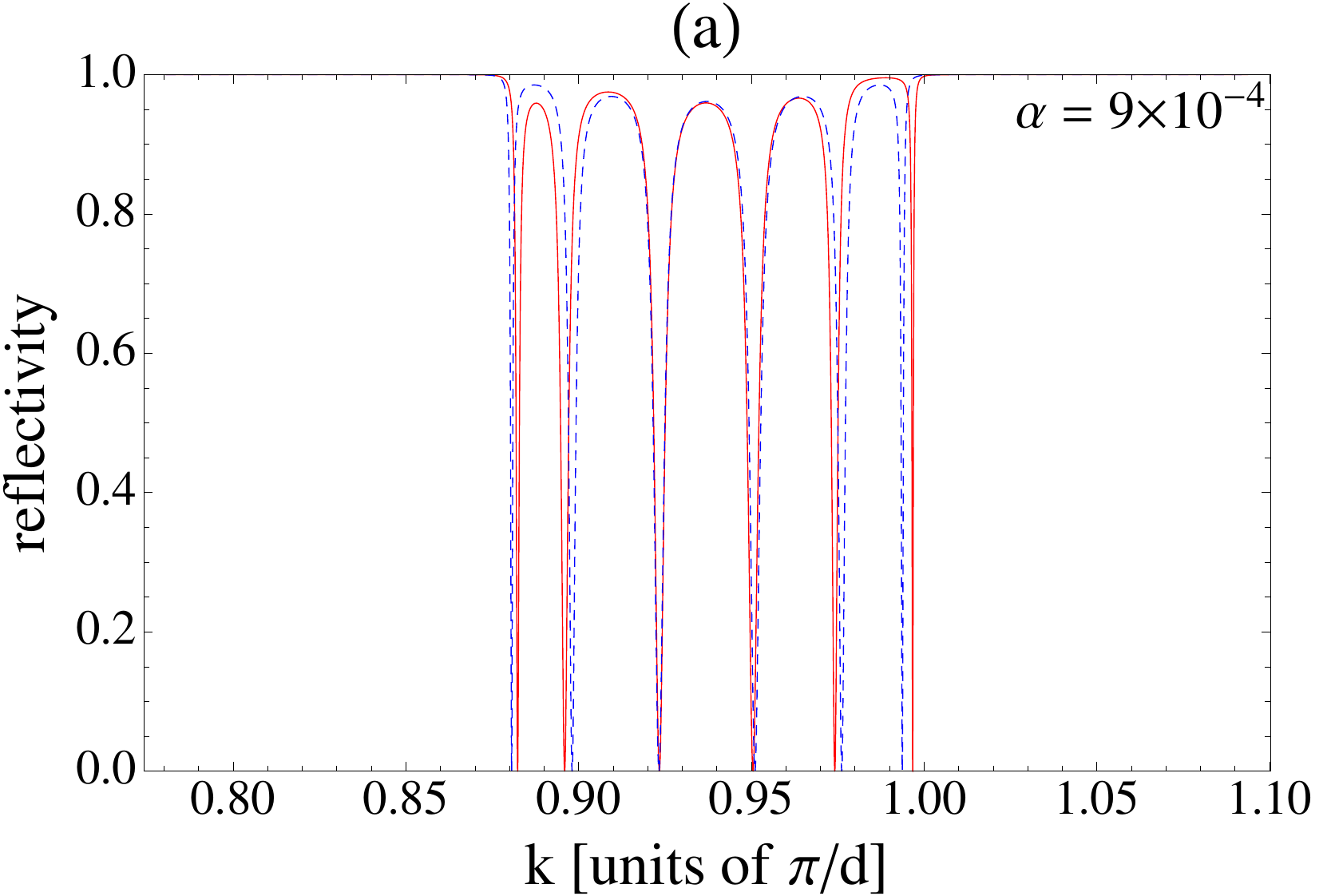}
\includegraphics[width=0.80\columnwidth]{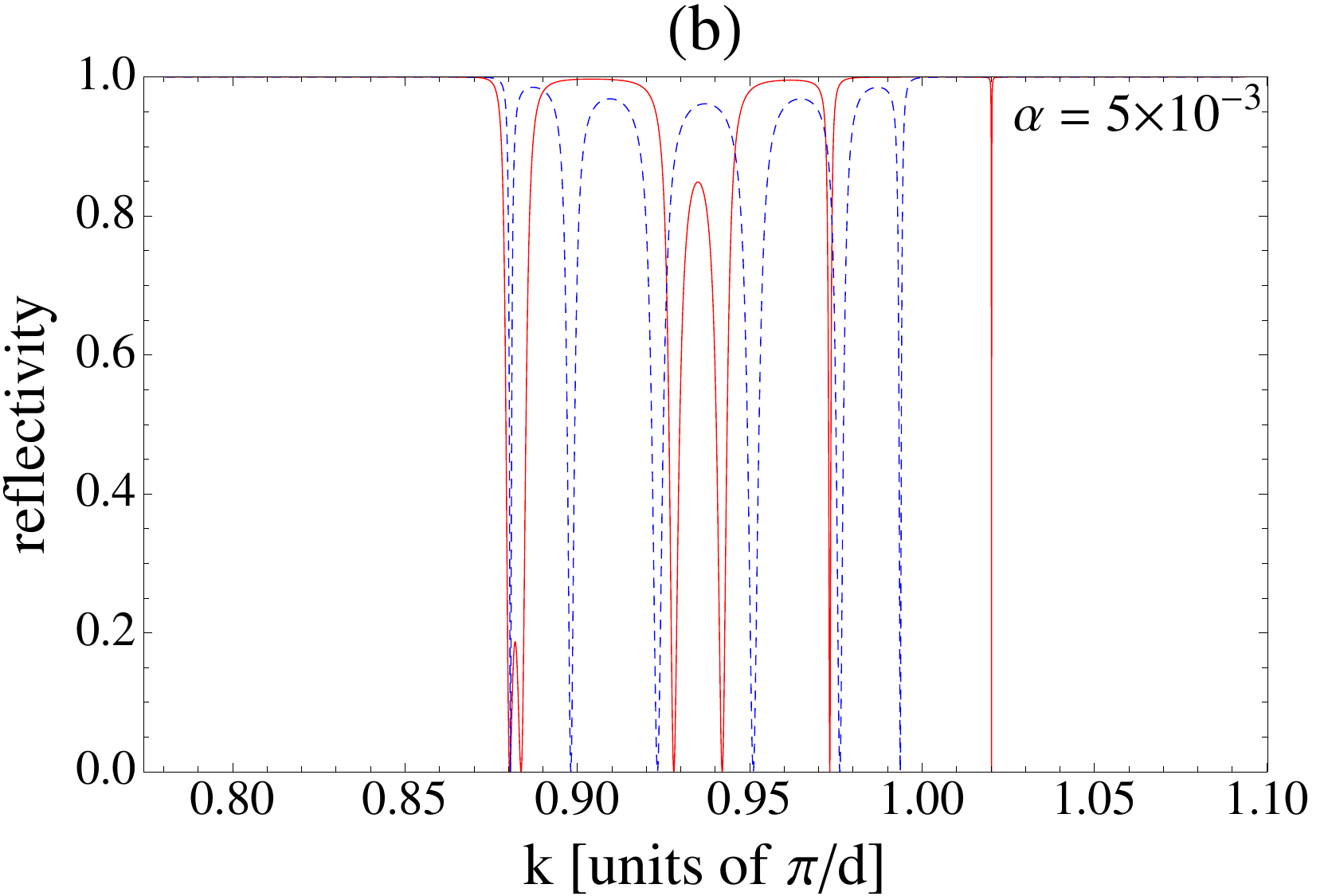}
\includegraphics[width=0.80\columnwidth]{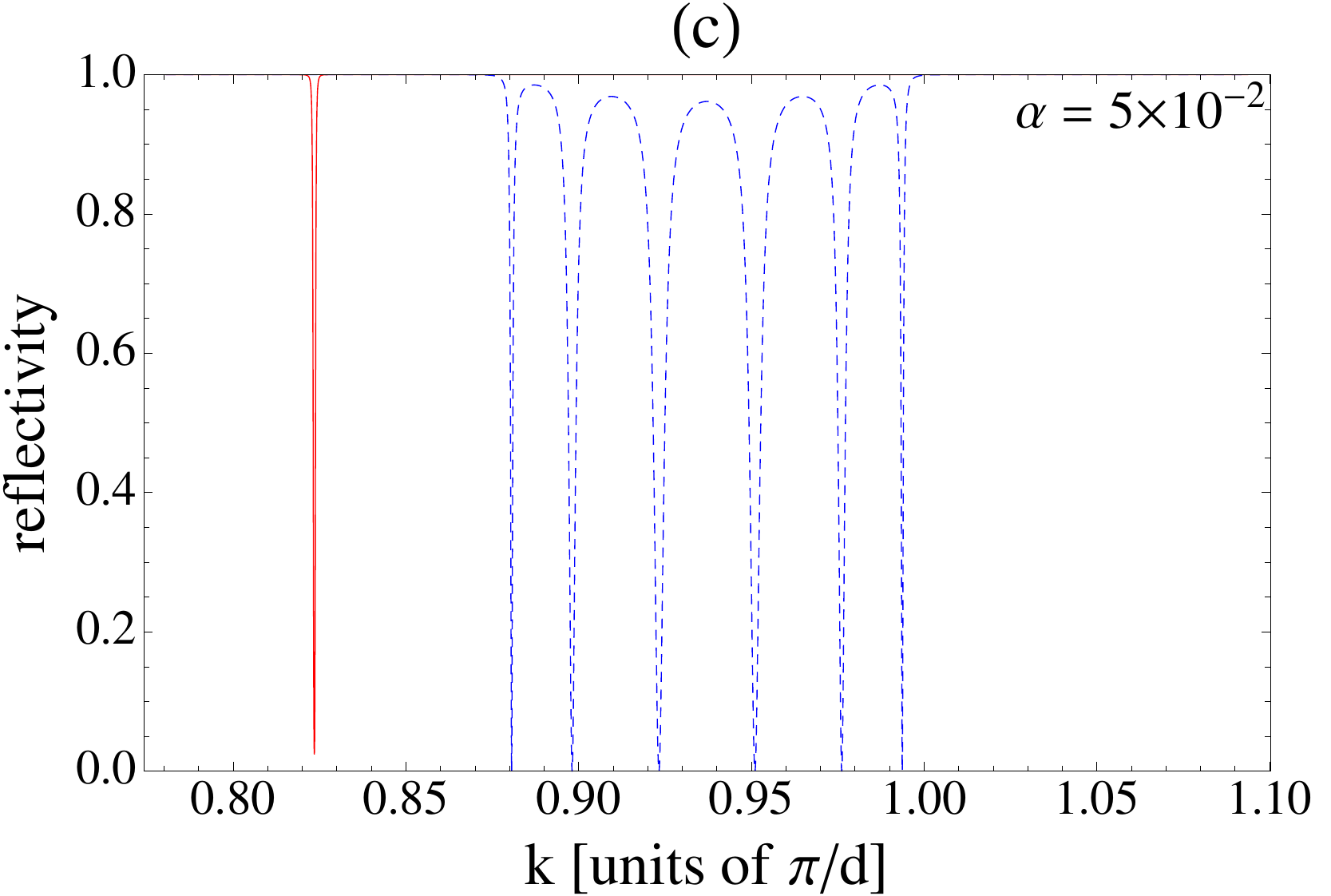}
\caption{\emph{Optical response}. Reflectivity of an array of seven
immobile membranes (each with $\zeta=-5$); from up to down the
figures compare the behavior of the array with defect (red continuous curves)
when $\alpha$ is scanned through values $9\times 10^{-4}$, $5\times
10^{-3}$, and $5\times 10^{-2}$ with the equidistant array (blue
dashed curves). For large $\alpha$, single resonances are singled
out inside the reflection band gap.}
\label{fig:fig2}
\end{figure}

Let us consider an equidistant membrane array inside a long, high-finesse cavity, as illustrated in Fig.~\ref{fig:fig1}a. We now recall some of the results of Refs.~\cite{xue12,xue13}, which show that optimal working points, that is choices of parameters that maximize the strength of the OM couplings $g^{(1)}_j$ and $g^{(2)}_j$,  are those where the membrane configuration ensures transparency, i.e.~around the zeros of the reflectivity function. For $N$ membranes there is an infinite set of {\it transmissive bands} each of them containing $N-1$ transmissive points (see App.~\ref{sec:appE} for a comparison with the band structure that occurs in the continuous limit where $N\rightarrow\infty$).  In the following we focus on the lowest-energy band and in particular to the first transmissive point where, as shown in Ref.~\cite{xue12}, the strength of the linear coupling $g^{(1)}_j$ for each individual membrane $j$ in the setup of Fig.~\ref{fig:fig1}a [see Eq.~\eqref{eq:eqCoup}] reflects the sinusoidal profile of the light field in the array as
\begin{equation}
g_j^{(1)} = G \sin\biggl(2\pi\frac{j-\tfrac{1}{2}}{N}\biggr).
\end{equation}
We have made the notation
\begin{equation}
G =-2 \omega_c x_0 \frac{\zeta\csc\bigl(\frac{\pi}{N}\bigr)\Bigl[\sqrt{\sin^2\bigl(\frac{\pi}{N}\bigr)+\zeta^2}-\zeta\Bigr]}{L-2Nd\zeta\csc^2\bigl(\frac{\pi}{N}\bigr)\sqrt{\sin^2\bigl(\frac{\pi}{N}\bigr)+\zeta^2}},
\end{equation}
and denoted the main resonance frequency of the cavity by $\omega_\mathrm{c}$, where $\omega_\mathrm{c}=c k$ (for the particular example shown as a blue (dashed) line in Fig.~\ref{fig:fig2}a, one has $\omega_\mathrm{c}\approx 0.87\pi c/d$). Note that the result holds more generally for any transmissive point, with some small modifications referring to the periodicity of the sine function (see Ref.~\cite{xue13}).

In terms of {\it collective vibrations} of the membrane array, the strength of the coupling of the lowest-energy collective mode is defined as
\begin{equation}
\label{eq:collcoup}
g^{(1)}_{\text{sin}}=\sqrt{\sum_{j=1}^{N-1}{g_{j}^{(1)}}^2}.
\end{equation}
For a small ratio $d/L$ and $N|\zeta|/\pi\ll 1$, one can show that this effective coupling reduces to $g^{(1)}_{\text{sin}}=g|\zeta|\sqrt{N/2}$. The $\sqrt{N}$-scaling is indeed typical of systems involving large ensembles of low reflectivity scatterers, as in atom-cavity OM~\cite{kru03,mur08}, where the sine mode reduces to an overall equal coupling mode, i.e., the center-of-mass mode. For $N|\zeta|/\pi\gg 1$ instead, as in our model, we obtain
\begin{equation}
g^{(1)}_{\text{sin}}\approx\frac{\sqrt{2}}{\pi}g\zeta^2N^{3/2},
\label{eq:g1simple}
\end{equation}
which can be orders of magnitude larger than $g$. In Sec.~\ref{subsec:refb} we show that the introduction of a spatial defect can further increase the achievable OM couplings. We also show an enhancement of the effective quadratic OM coupling, where a figure of merit is defined as
\begin{equation}
\label{eq:collcoup}
g^{(2)}_{\text{sin}}=\sqrt{\sum_{j=1}^{N-1}{g_{j}^{(2)}}^2}.
\end{equation}

\subsection{Array with quadratic spatial defect}
\label{subsec:IIIB}
Let us now depart from the equidistant case by considering the spatial positioning defined in Eq.~(\ref{eq:quadpos}). Since $M_i=M_j$ for any $i, j\in\{1,...,N\}$, we can drop all indexes. The transfer matrix of the whole array is
\begin{equation}
\label{eq:ar}
M_{ar}=M\cdot\left(\prod_{j=1}^{N-1}F_{j,j+1}\cdot M\right)
\end{equation}
where $M=\mathbb{I}+i\zeta(\sigma_1+i\sigma_2)$ and $F_{j,j+1}=e^{ik(x_{j+1}-x_j)\sigma_3}$ are written in terms of the Pauli matrices $\sigma_1, \sigma_2, \sigma_3$ and identity matrix $\mathbb{I}$.
The correction to the equidistant case is
\begin{equation}
\mathrm{d}_{j,j+1}=\frac{d-\left(x_{j+1}-x_j\right)}{\alpha},
\end{equation}
so that $d_{j,j+1}=d-\alpha \mathrm{d}_{j,j+1}$. Owing to defect symmetry with respect to reflection about the origin, the correction satisfies $\mathrm{d}_{i,i+1}=\mathrm{d}_{N-i,N-i+1}$ for any positive integer $i$ belonging to the set $\{1,[(N-1)/2]\}$ (we defined $[(N-1)/2]\equiv\text{Floor}[(N-1)/2]$). For small $\alpha$, the defect introduces a perturbation to the equidistant case, which allows one to apply the MacLaurin expansion up to order $O(\alpha^2)$
\begin{equation}
\label{eq:Fj}
F_{j,j+1}=(\mathbb{I}-ik\alpha \mathrm{d}_{j,j+1}\sigma_3) \cdot F + O(\alpha^2).
\end{equation}
Matrix $F$ describes the propagation of a monochromatic beam with wave number $k$ over a distance $d$ (i.e., with $\alpha=0$), $F=e^{ikd\sigma_3}$.
Inserting Eq.~\eqref{eq:Fj} in Eq.~(\ref{eq:ar}) and collecting
terms proportional to $\alpha$ one can rewrite the transfer
matrix as
\begin{equation}
\label{eq:Mtot}
M_{ar}=M_{N}+\alpha M_{corr}+O(\alpha^2),
\end{equation}
with more details of the derivation presented in Appendix \ref{sec:appA}.
In the absence of any defects ($\alpha=0$), $M_{ar}$ reduces to $M_{N}$, which, as shown in Refs.~\cite{xue13}, can be recast in the form
\begin{equation}
\label{eq:paraform}
M_{N}=\begin{bmatrix} (1+i\chi)e^{i\mu}& i\chi \\ -i\chi & (1-i\chi)e^{-i\mu} \end{bmatrix},
\end{equation}
with an effective polarizability $\chi=\zeta U_{N-1}(a)$ and effective phase $\mu$ obeying
\begin{equation}
\label{eq:mu}
e^{i\mu}=\frac{1-\zeta U_{N-1}(a)}{(1-i\zeta)U_{N-1}(a)-e^{ikd}U_{N-2}(a)}.
\end{equation}
Here $a\equiv a(kd)=\cos(kd)-\zeta\sin(kd)$, [see Eq. \eqref{eq:a}], and $U_j$ is the Chebyshev polynomial of the second kind of degree $j$.
The term proportional to $\alpha$ in Eq.~\eqref{eq:Mtot} can be rewritten as
\begin{equation}
\label{eq:Mcorr}
M_{corr}=\begin{bmatrix} (1+i\xi)e^{i\nu}& i\xi \\ -i\xi & (1-i\xi)e^{-i\nu} \end{bmatrix},
\end{equation}
with first-order polarizability $\xi$ defined by
\begin{equation}
\label{eq:xi}
\xi=4\zeta kb \sum_{j=1}^{[N/2]}\mathrm{d}_{j,j+1}\left(1-\frac{\delta_{j,N/2}}{2}\right)U_{j-1}(a)U_{N-j-1}(a),
\end{equation}
where $\delta_{j,N/2}$ is a Kronecker delta, and first-order phase $\nu$ defined by
\begin{equation}
\label{eq:nu}
e^{i\nu}=\frac{(i+\xi)e^{-ikd}}{k\sum_{j=1}^{[N/2]}\mathrm{d}_{j,j+1}\left(\frac{\delta_{j,N/2}}{2}-1\right)[C]_{22}}.
\end{equation}
Function $b$ in Eq.~(\ref{eq:xi}) corresponds to $a$ after an argument shift $kd\rightarrow kd-\pi/2$, [see Eq.~\eqref{eq:b}]. Function $[C]_{22}$ is given by the second line of Eq.~(\ref{eq:C}).

\begin{figure}[t]
\includegraphics[scale=0.45]{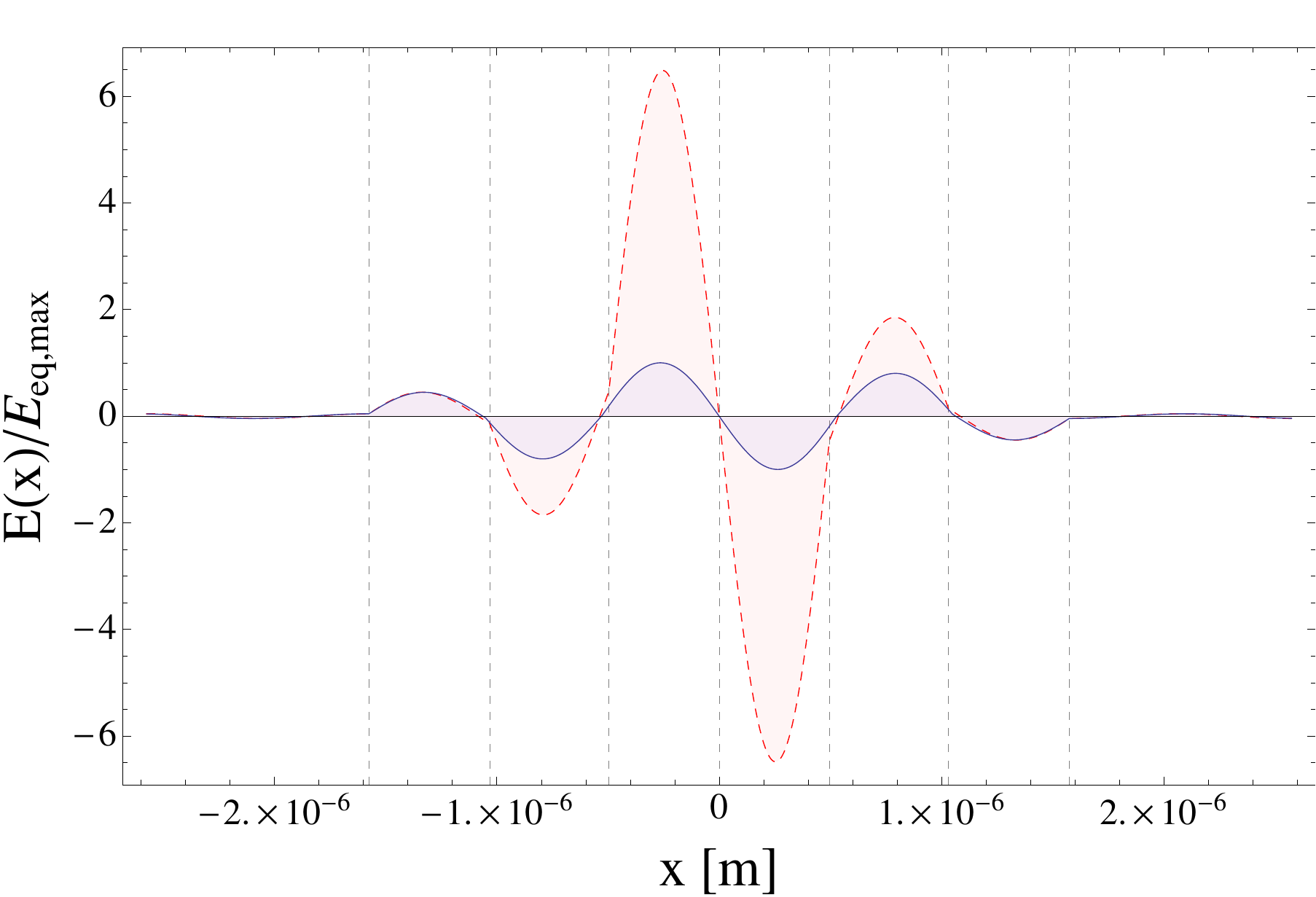}
\caption{\emph{Field localization}. Plot of the electric field
amplitude normalized to its equidistant case values along the array (vertical dashed lines show the membrane positions) for two situations in both of
which the system is transparent: (i) no defect ($\alpha=0$) (blue,
solid line) and (ii) with defect ($\alpha=5\times10^{-3}$) (red, dashed line). While the
equidistant array already shows localization of the field mode, the
defect can enhance this effect. The field amplitude is normalized to the maximum value achieved for the equidistant case.}
\label{fig:fig3}
\end{figure}

To first order in $\alpha$ we have therefore
\begin{equation}
\label{eq:Mtot1}
M_{ar}=\begin{bmatrix} (1+i\gamma)e^{i\lambda}& i\gamma \\ -i\gamma & (1-i\gamma)e^{-i\lambda} \end{bmatrix},
\end{equation}
where the effective polarizability $\gamma$ and the effective phase $\lambda$ are given by
\begin{empheq}[left=\empheqlbrace]{align}
\label{eq:graf}
&\gamma=\chi+\alpha\xi,\\
\label{eq:lamb}
&e^{i\lambda}=\frac{1-i(\chi+\alpha\xi)}{(1-i\chi)e^{-i\mu}+\alpha(1-i\xi)e^{-i\nu}}.
\end{empheq}
Notice that, for a vanishing defect $\alpha=0$, $\gamma$ and $\lambda$ reduce to $\chi$ and $\mu$ respectively, as expected.

However, the validity of the above first order expansion is restricted to sufficiently small values of $\alpha$ (for example $\alpha<10^{-3}$) (for which the optical response is plotted in Fig.~\ref{fig:fig2}a). For increasing values of $\alpha$, where much stronger modifications of the optical response occur, as illustrated in Fig.~\ref{fig:fig2}b,c, we mainly use numerical tools for deriving the OM couplings.

\subsection{Numerical results}
\label{subsec:refb}

In Fig.~\ref{fig:fig2} the optical response [reflectivity as a function of the wave vector $k$ of the incoming electric field $E(x,t)$] of the free-standing membrane array in the presence of a defect (continuous red line) is compared to the equidistant case ($\alpha = 0$, dashed blue line), for $\alpha=9\times 10^{-4}, 5\times 10^{-3}$, and $5\times 10^{-2}$ (up to down). We chose $N=7$, $\zeta=-5$, and the separation in the absence of the defect is $d=525$ nm.  While both cases $\alpha=0$ and $\alpha\neq 0$ display an infinite number of bands, here we focus on the first band with $k>0$ only.

The figure shows that for a small defect strength $\alpha=9\times 10^{-4}$ the reflectivity is very similar to that of the equidistant case, where all resonances are confined to a well-defined band of width $2\arcsin[\cos(\pi/N)/\sqrt{1+\zeta^2}]$. However, for increasing $\alpha$ we observe a shift of the position of the resonances, as well as a redistribution of their degeneracies. For example, in Fig.~\ref{fig:fig2} a doubly-degenerate resonance is seen to shift towards larger values of $k$ (i.e., $kd/\pi \sim 1.02$), while in Fig.~\ref{fig:fig2}c two degenerate resonances appear at low values of $kd/\pi\sim 0.825$, within the band gap.

Because of the quadratic character of the chosen defect (see App.~\ref{sec:appendixC}), here the wave-functions of the modes resemble modified Hermite polynomials. This allows one to engineer larger gradients of the electric field across individual membranes, compared to the plane-wave behavior of the case $\alpha=0$. This is important since, as explained in Sec.~\ref{sec:toa} above, the linear coupling $|g^{(1)}_j|$ at the membrane $j$ is directly proportional to the local field gradient (while its sign depends on whether the maximal amplitude of the field is on the left or right of the membrane). This enhancement is exemplified in Fig.~\ref{fig:fig3}, where the amplitude $E(x)$ of the electric field is plotted as a function of $x$ for the case of the higher-energy resonance in Fig.~\ref{fig:fig2}b, with $kd/\pi\sim 1.02$. We find that this kind of higher-$k$ resonance is in fact the most favorable for obtaining large couplings with the quadratic defects considered here, as also discussed below. We note that, for sufficiently large $\alpha$, resonances can disappear due to destructive interference effects. This is the case in Fig.~\ref{fig:fig2}c, where all but two of the transmissive resonances have disappeared.


In the following, we consider the compound system of the array discussed above and the surrounding optical cavity. We investigate numerically the OM linear and quadratic couplings in the vicinity of common transparency points of the array and of the cavity (chosen length $L=6.3\,$cm). The two end mirrors of the cavity have a polarizability $\zeta_m=-20$.

\begin{figure}[t]
\centering
\includegraphics[width=0.93\columnwidth]{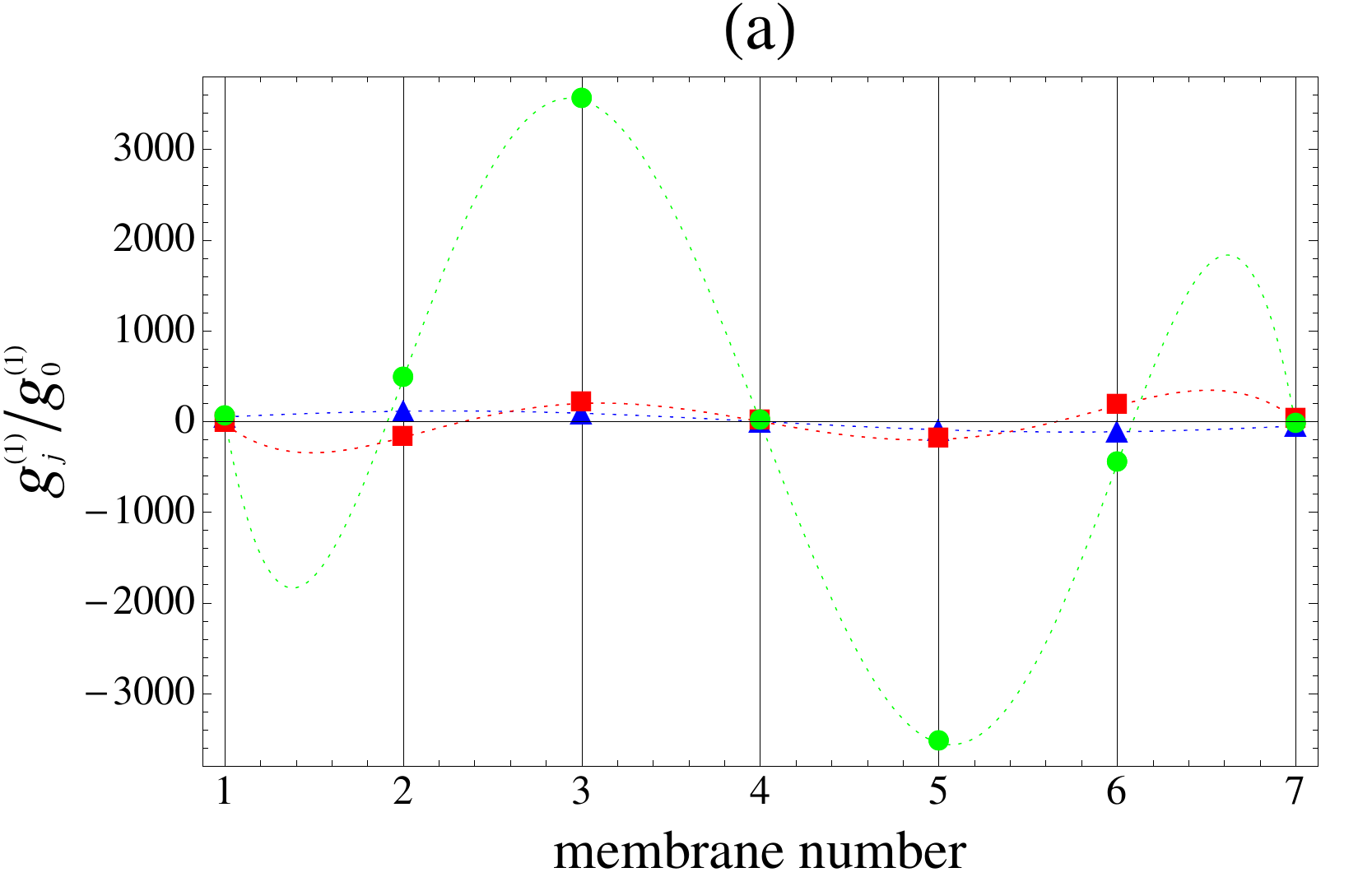}
\includegraphics[width=0.93\columnwidth]{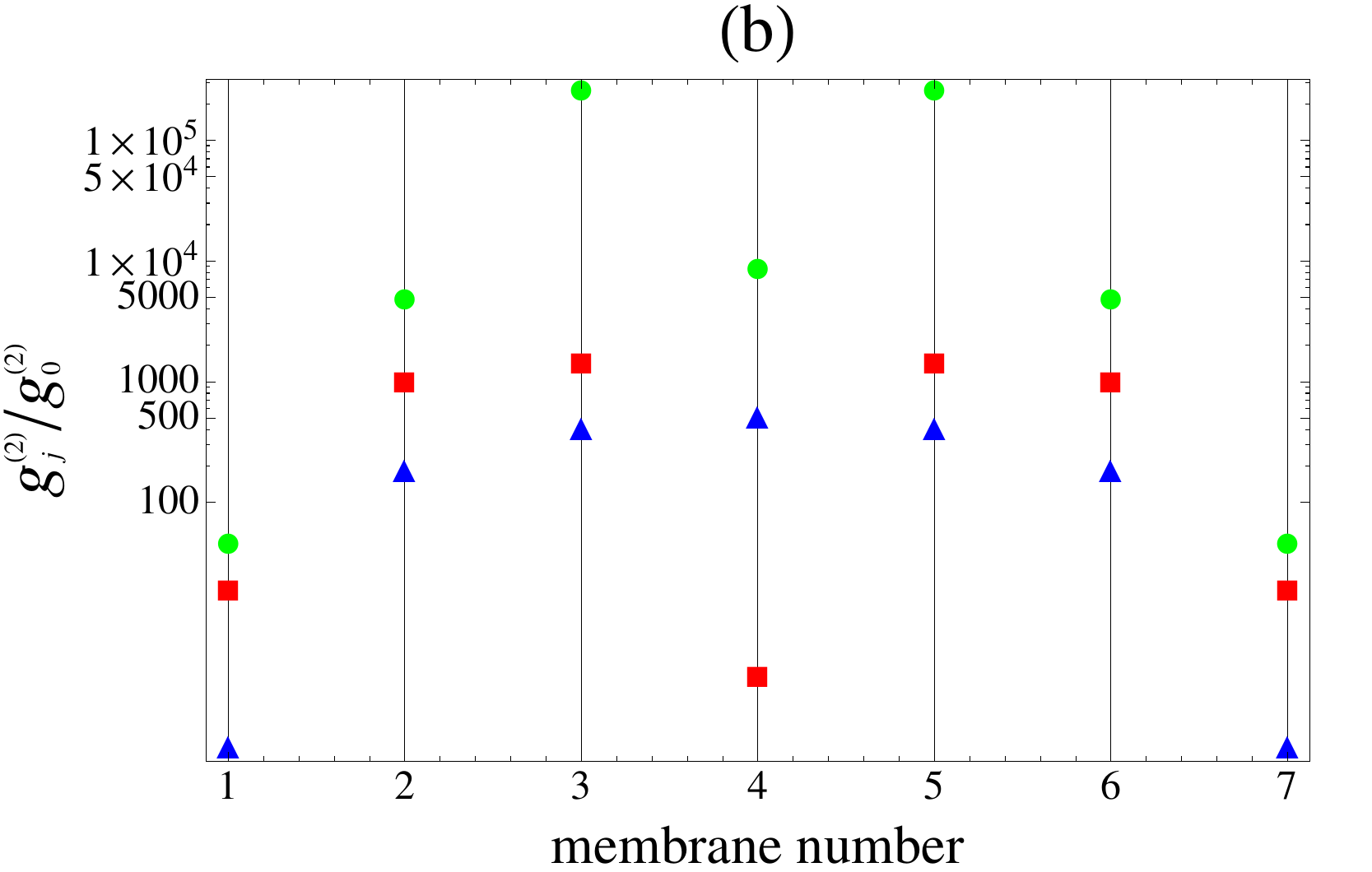}
\caption{\emph{OM coupling strengths}. Behavior of linear (a) and quadratic (b) OM couplings
for an array of 7 membranes in a cavity of length $L=6.3\times 10^{-2}$ m with an inter-membrane distance $d=525$ nm. The polarizability of the cavity mirrors is $\zeta_{m}=-20$ and every membrane has a polarizability $\zeta=-5$. The three thin curves blue/green/red, correspond to $\alpha=0$, and $\alpha=5\times 10^{-3}$ and $\alpha= 5\times 10^{-2}$ respectively. For the numerical example considered, the introduction of the defect builds on the enhancement provided already by the access of the last transmission point by increasing $g^{(1)}/g_0^{(1)}$ by a factor of $23$  and $g^{(2)}/g_0^{(2)}$ by a factor $434$. Notice that we fixed $g^{(1)}/g_0^{(1)}$ for the equidistant case to the maximum value allowed, roughly equal to: $\sqrt{2}/\pi\zeta^2 N^{3/2}\simeq217$. For an even larger defect ($\alpha=5\times 10^{-2}$), only one resonance survives and it is moved into the lower energy band gap instead, with corresponding lower enhancement factors $1.73$ and $2.94$.}
\label{fig:fig4}
\end{figure}

The main results are illustrated in Fig.~\ref{fig:fig4} for a few resonances corresponding to the different situations depicted in Fig.~\ref{fig:fig2}. In particular, the blue triangles correspond to the couplings for the resonances with the leftmost $k$ of Fig.~\ref{fig:fig2}a (with $\alpha=0$), the green dots to the rightmost $k$ of Fig.~\ref{fig:fig2}b and the red squares to the leftmost $k$ in Fig.~\ref{fig:fig2}c (with $\alpha= 5 \times 10^{-2}$). The figure shows that the presence of the defect can lead in general to orders-of-magnitude enhancement of both linear and quadratic OM couplings with respect to the case with $\alpha=0$.

The reference blue curve for $\alpha=0$ in Fig.~\ref{fig:fig4}a, fits the expected analytical results showing an enhancement of about $217$ consistent with the expected scaling $g^{(1)}_{\text{sin}}/g^{(1)}_0\approx \sqrt{2}/\pi\zeta^2 N^{3/2}$.
The reference effective quadratic coupling (blue curve in Fig.~\ref{fig:fig4}b) reaches a value of $0.789\times10^3$ relative to the single element quadratic coupling $g_0^{(2)}$. The relatively small defect of $\alpha=5\times10^{-3}$ shifts the rightmost $(N-1)$-th resonance into the first gap, as illustrated in Fig.~\ref{fig:fig2}b. Large enhancement of both linear and quadratic OM couplings by factors of about $23$ and $430$ over the equidistant case occur. Further increase of the defect reduces the number of available resonances to a single one pushed inside the lowest energy band gap with corresponding lower enhancement factors $1.7$ and $2.9$.
While above we have described the relative improvement brought on by the quadratic defect, we provide now a more explicit experimental case study. Let us consider membranes with frequency $\omega_m=2\pi\times211$ kHz and zero-point motion $x_0=2.7$ fm. For the single membrane OM one would then obtain an optimal coupling $g=2\pi\times24$ Hz which is small compared to both $\omega_m$ and realistic cavity decay rates (typically in the range of $0.1\div1$ Mhz). The enhancement brought about by addressing collective modes of the equidistant array of $7$ membranes each with $\zeta=-5$ brings already the coupling to $2\pi\times 5$ kHz. The introduction of the defect leads to a total coupling $2 \pi\times117$ kHz already comparable to typical mechanical resonant frequencies and optical cavity decay rates. For quadratic coupling, the realistic single membrane values are extremely small around $2\pi\times2\times10^{-6}$ Hz. The enhancement from operating at transmissive points combined with the extra-localization induced by the defect can bring this value to $2\pi\times1.4$ Hz, rendering it potentially observable in realistic OM experiments concerned with direct optical monitoring of quantum jumps in phonon numbers of mechanical resonators.

\section{Optomechanical crystals}
\label{sec:secIV} The 1D transfer matrix approach can provide the basis for both analytical and numerical studies of phonon-photon couplings on the OM crystal platform. As already stated in Sec.~\ref{sec:toa}, we model an OM crystal as a device composed of three well-defined parts: two equidistant and periodic side arrays (forming effective left/right extended mirrors) and one quasi-periodic array in the middle
[see device \textbf{b} in Fig.~\ref{fig:fig1}]. The two extended mirrors
form a ``super-cavity" surrounding the quasi-periodic array whose
optical properties can be analytically investigated by making use of
results on the equidistant arrays from the previous section. We then
insert the array in the presence of the defect $\alpha$ inside the super-cavity (see Fig.~\ref{fig:fig6}c) and estimate the OM couplings.

\begin{figure}[t]
\includegraphics[width=0.99\columnwidth]{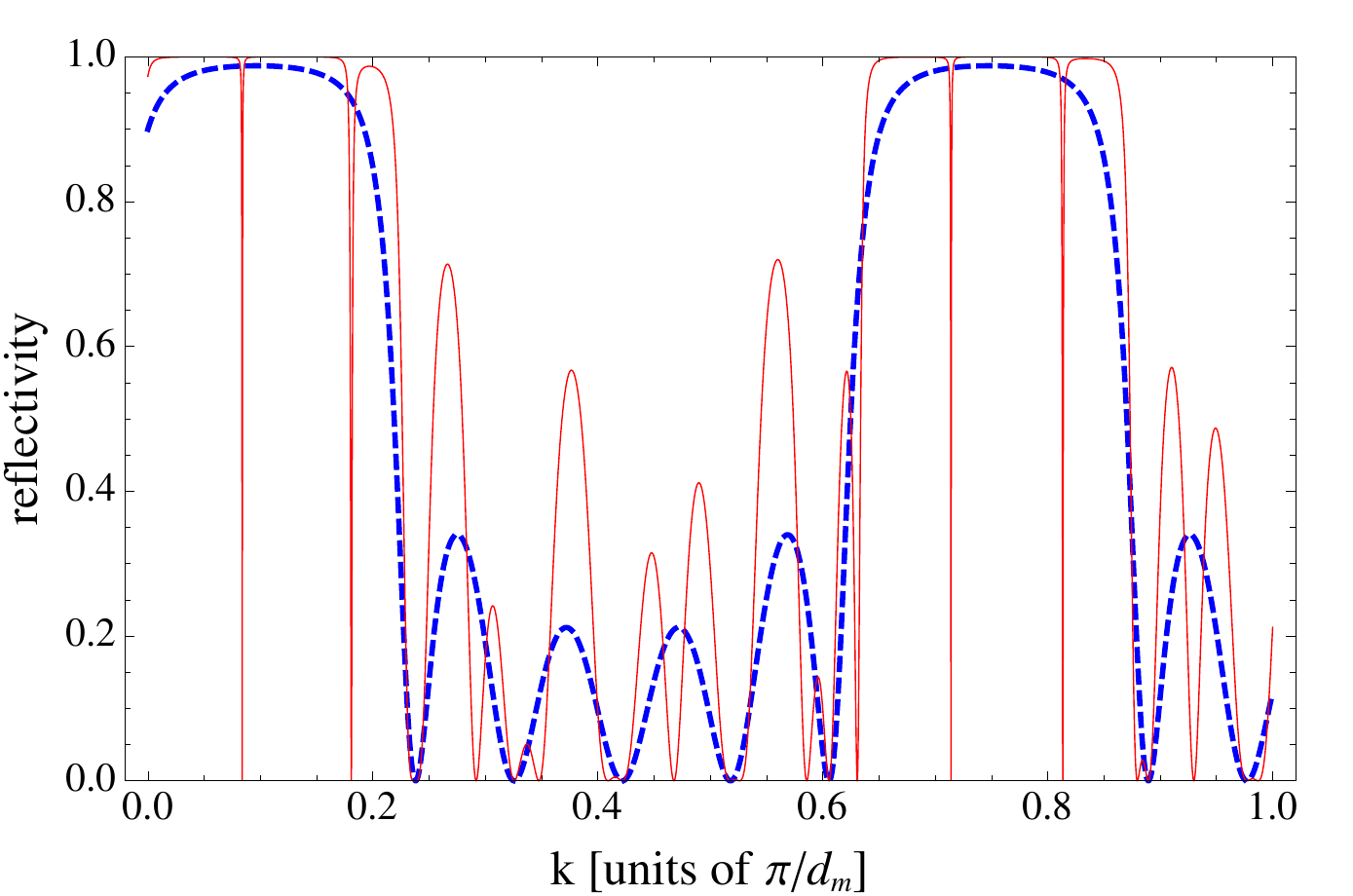}
\caption{Optical response of an empty super-cavity (continuous red curve) overlapped to that of a single side mirror (dashed blue curve) for $N_m=6$ and $\zeta_m=-0.5$, and $d_m=768$ nm.}
\label{fig:fig5}
\end{figure}

\subsection{Super-cavity}
\label{subsec:subsecIVA}
We treat the side periodic arrays of the OM crystal as $k$-dependent mirrors with an effective
polarizability $\chi_m=\zeta_m U_{N_m-1}(a_m)$ and a phase $\mu_m$ defined by Eq.~(\ref{eq:mu}) with polarizability $\zeta_m$, membrane number $N_m$ and separation $d_m$. These two mirrors form the super-cavity. Assuming that the super-cavity is empty, one can readily compute its transmission function as
\begin{equation}
\label{eq:trec}
T=\frac{1}{|e^{-i(kD+2\mu_m)}(1-i\chi_m)^2+e^{ikD}\chi_m^2|^2}.
\end{equation}
In Fig.~{\ref{fig:fig5} the reflectivity $1-T$ is plotted (red continuous curve) as a function of the wave vector of the incoming electric field (in units of $\pi/d_m$). The blue (dashed) curve instead illustrates the reflectivity of a single side mirror.

We note that $1-T$ has two different types of resonances: (i) formed by
the overlapping of common resonances of the side mirrors and thus lying in a band for the reflectivity of a single side mirror (e.g. red transmission points at $kd/\pi\approx 0.29,\,0.39$ up to $0.6$ in Fig.~{\ref{fig:fig5}), (ii) obtained in
the regime where both side mirrors have a reflectivity close to unity. We are interested in the latter ones, which lie within a band gap for the reflectivity of a single side mirror and describe a high-finesse optical cavity. Examples of these resonances are red transmission points at $kd/\pi\approx 0.71$ or $0.81$ in Fig.~{\ref{fig:fig5}. In such a regime, the dispersion curve of the side mirrors is practically flat and the super-cavity is well defined by the mirror polarizability $\chi_m$ and length $D$. This situation is illustrated in Fig.~{\ref{fig:fig5}} for $N_m=6$; the linewidth of the peaks is inversely proportional to $\chi_m^2=N_m^2U_{N_m-1}(a_m)^2$.

\begin{figure}[b]
\includegraphics[width=0.99\columnwidth]{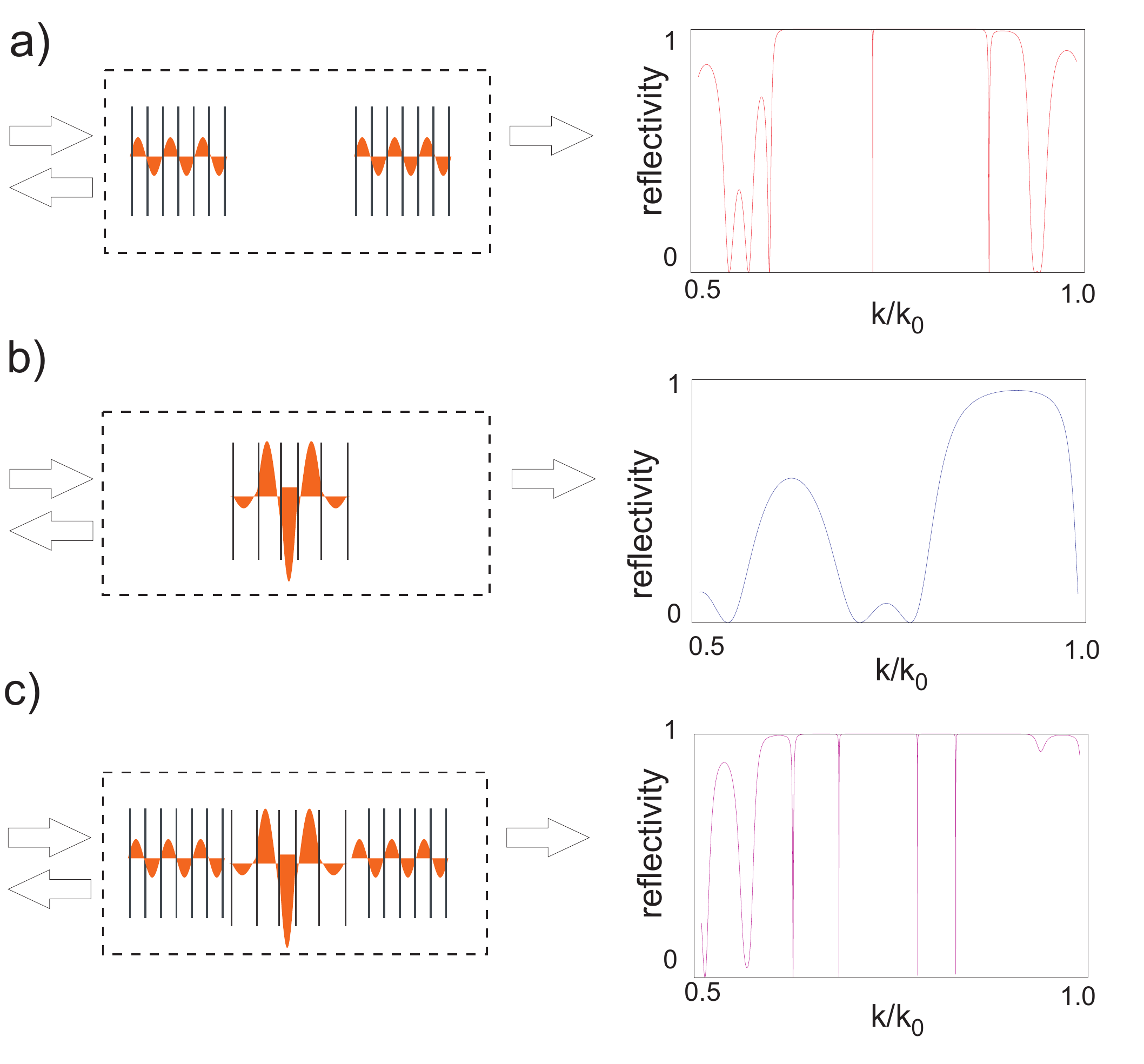}
\caption{\emph{OM crystal optical response} (a)
An empty super-cavity shows resonances in the common band gaps of the individual extended side mirrors; in fact the plot displays the reflectivity of the super-cavity as a function  the wave vector k (in units of $k_0=\pi/d$). (b) A membrane array with engineered quadratic defect can show, under certain conditions, field localization. (c) Super-cavity with quasi-periodic array inside simulates an OM crystal. The optical response is a convolution of the two previously plotted responses. To find transmissive regimes one finds the wave vectors at which both upper and middle reflectivity plots show zeros.}
\label{fig:fig6}
\end{figure}

\noindent An estimate of the finesse shows that
\begin{equation}
\label{eq:Fin}
F=\frac{k_{\text{FSR}}}{\kappa_{k}}=\frac{\pi}{D}\frac{D|\chi_m|\sqrt{1+\chi_m^2}}{\pi}=|\chi_m|\sqrt{1+\chi_m^2}\approx\chi_m^2,
\end{equation}
showing large values of the finesse for large values of $\chi_m$. Equation.~\eqref{eq:Fin} stems from the fact that, for a Fabry-P\'erot resonator with length $D$ and side-mirrors polarizability $\chi_m$, the free spectral range in $k$ is given by $k_{\text{FSR}}=\pi/L$ while the linewidth is
\begin{equation}
\kappa=\frac{\pi}{L|\chi_m|\sqrt{1+\chi_m^2}}.
\end{equation}

In this limit, $\chi_m\gg1$, we can approximate the total transmission of Eq.~(\ref{eq:trec}) by
\begin{equation}
T\simeq\frac{1}{\chi_m^4|1-e^{-2i(kD+\mu_m)}|}.
\end{equation}
The condition for resonance ($T=1$) reduces to
\begin{equation}
\label{eq:condforres}
|1-e^{-2i(kD+\mu_m)}|=\frac{1}{\chi_m^4}
\end{equation}
which in view of $\chi_m\gg1$ forces the left side of Eq.~\eqref{eq:condforres} to zero. We obtain then
\begin{equation}
\label{eq:fond}
k^{(n)}D+\mu_m=n\pi\;\;\;\text{with}\;\;\;n\in\mathbb{Z}.
\end{equation}
We note that $k^{(0)}=0$ because $\mu_m$ is exactly zero at $k=0$. This can be easily seen from Eq.~(\ref{eq:mu}) by using the fact that $a=1$ at $k=0$ and $U_n(1)=n+1$ for every $n$. The first positive resonance is then $k^{(1)}$, the second $k^{(2)}$, and so on.
After some algebraic passages, and having employed Eq.~(\ref{eq:mu}), Eq.~(\ref{eq:fond}) can be rewritten as:
\begin{equation}
\label{eq:caso1}
\frac{U_{N_m-2}(a_m^{(n)})}{U_{N_m-1}(a_m^{(n)})}=\left[1-i\zeta_m(1-e^{i(k^{(n)}D-n\pi)})\right]e^{-ik^{(n)}d_m}
\end{equation}
where $a_m^{(n)}$ is simply $a_m$ evaluated at $k=k^{(n)}$. One can simplify the expression even more because
\begin{equation}
\label{eq:caso2}
\begin{split}
\frac{U_{N_m-2}(a_m^{(n)})}{U_{N_m-1}(a_m^{(n)})}&=\frac{\sin((N_m-1)\arccos(a_m^{(n)}))}{\sin(N_m\arccos(a_m^{(n)}))}\\
&=a_m^{(n)}-\sqrt{1-a_m^{(n)2}}\cot(N_m\arccos(a_m^{(n)})),
\end{split}
\end{equation}
and for a given purely-imaginary number $z$ such that $|z|\gg0$, it holds true that $\cot z \rightarrow -i$. Applying this to Eqs.~(\ref{eq:caso1}) and (\ref{eq:caso2}) we obtain
\begin{equation}
\tan(k^{(n)}d_m)=   \zeta_m[(-1)^n\cos(k^{(n)}(d_m-D))-1].
\end{equation}
The expression above provides a simpler equation for the resonant wave vectors, valid under the assumptions:
\begin{equation}
|\arccos(a_m^{(n)})|\ll \frac{1}{N_m}\;\;\;\text{and}\;\;\;|\zeta_m|>\frac{1-\cos(kd_m)}{\sin(kd_m)}.
\end{equation}

We note that although the $k^{(n)}$'s are practically standard-cavity resonances, they are not separated by a free spectral range $\pi/D$. There is an extra factor $\Delta_{(i,j)}$ which depends on the distance $d_m$ between consecutive membranes in the same side mirror:
\begin{equation}
k^{(i)}-k^{(j)}=(i-j)\frac{\pi}{D}+\Delta_{(i,j)},
\end{equation}
with
\begin{equation}
\Delta_{(i,j)}=\frac{1}{iD}\log \left(\frac{1-\frac{e^{ik^{(i)}d_m}}{1-i\zeta_m}\frac{U_{N_m-2}(a_m^{(i)})}{U_{N_m-1}(a_m^{(i)})}}{1-\frac{e^{ik^{(j)}d_m}}{1-i\zeta_m}\frac{U_{N_m-2}(a_m^{(j)})}{U_{N_m-1}(a_m^{(j)})}}\right).
\end{equation}
To recover the standard Fabry-P\'erot resonator with dispersionless mirrors, we set $d_m=0$ and obtain $\Delta_{(i,j)}=0$ resulting in a constant free spectral range. This result allows one to use $d_m$ as a knob for tuning the position of resonances.

\subsection{Super-cavity with defect inside}
The transmissive regime is reached by simultaneously tuning the resonances of the super-cavity and the defect region. To this end we follow the steps illustrated in Fig.~{\ref{fig:fig6}} where we first identify the wide resonances of the defect area, and then find a close super-cavity resonance within the band gap which we tune by varying $d_m$ such that it coincides with a resonance of the defect area.
The OM crystal reduces to a couple of side mirrors composed of $N_m+1$ elements with inter-membrane separation $d_m$, and a defect region containing $N_d=N-2$ membranes. We fix in the following $\zeta_m=\zeta$ and illustrate the procedure of finding the resonances of the crystal in Fig.~\ref{fig:fig7} where the first 8 super-cavity resonant wave vectors (dotted lines) are tuned by adjusting $d_m$; We chose $N_m=20$, $N=7$ and $\zeta=-0.5$ ($20\%$ reflectivity membranes). Numerical values of the resonances are found by exploiting Eq.~(\ref{eq:fond}) with $n\in[1,8]$ which is valid only under the assumption that the empty cavity has a high finesse. All dotted lines present irregular breaches of ``fake" resonances which correspond to the transmission points in Fig.~\ref{fig:fig5} located between band gaps; these are indeed resonances of the empty cavity but with low finesse and Eq.~(\ref{eq:fond}) cannot be applied around these points. Horizontal dashed lines represent the resonant wave vectors of the quasi-periodic array and are here plotted for $N=7$ and $\alpha=10^{-3}$. Black squares are centered about the common resonances of the super-cavity and the array. Red triangles mark the corresponding overall crystal resonances which are slightly shifted from the squares owing to the shift introduced by $\mu_m$ [see Eq.~(\ref{eq:mu}) with $N_m\rightarrow N_m+1$]. We consider only resonances which, despite the shift, keep lying inside the super-cavity band gap.

\begin{figure}[t]
\includegraphics[width=0.89\columnwidth]{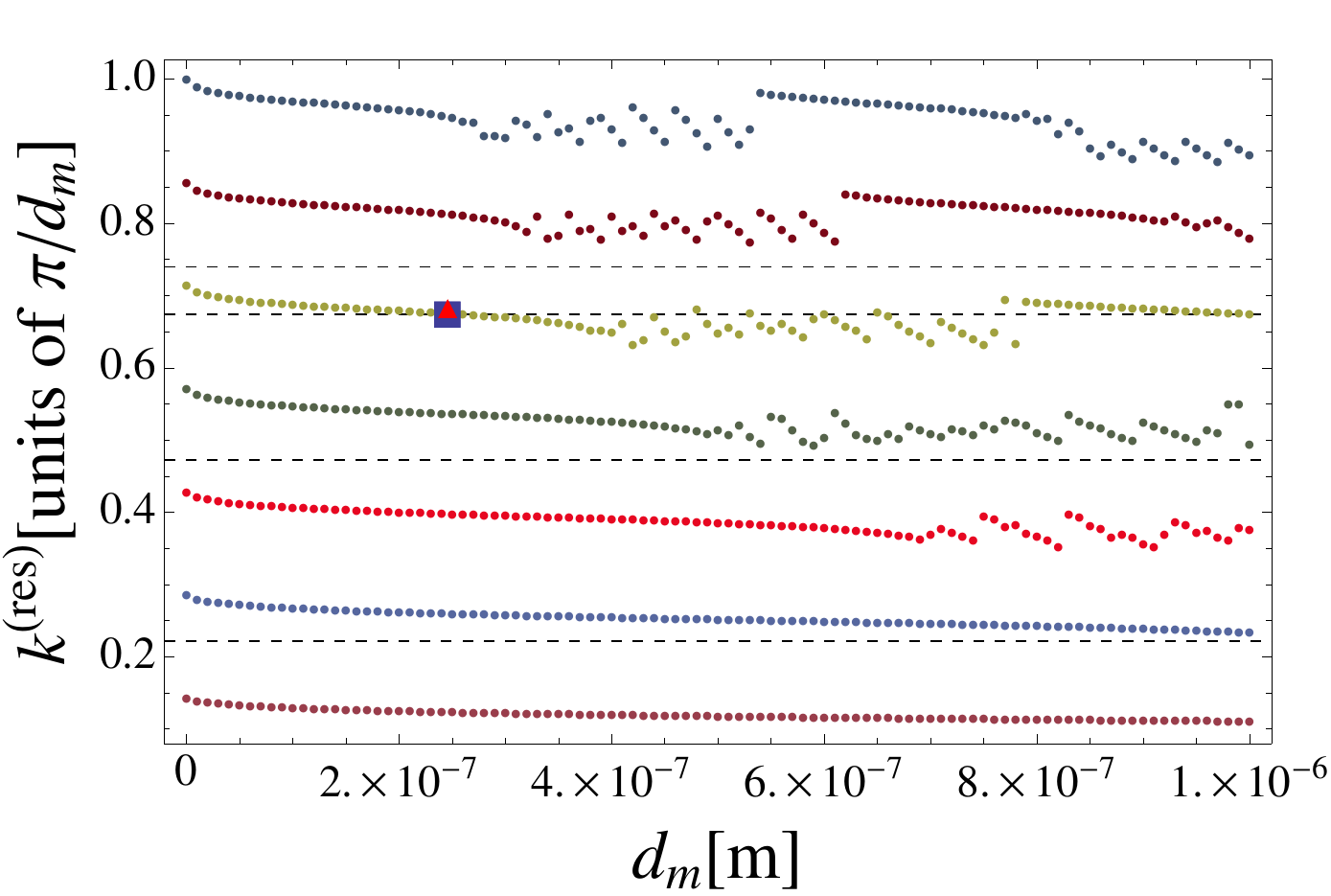}
\caption{\emph{Tuning of resonances} Common resonances are obtained as the intersections of the empty super-cavity resonances (dotted lines) with middle element resonances (dashed lines) as the inter-membrane distance $d_m$ is varied. The first 8 empty super-cavity resonances are plotted and the black square indicates a common resonance operating point. Notice that the scattered plot regions correspond to the empty-cavity resonances between the band gaps where the effective finesse is small and the analysis is not valid. The red triangle shows the small shift of the crystal resonances with respect to the common resonance and is due to the phase shift $\mu_m$. The side mirrors consist of $20$ elements while the middle region contains $7$ membranes.  We also chose $\alpha=10^{-3}$, and $\zeta=\zeta_m=-0.5$.}
\label{fig:fig7}
\end{figure}



Once the desired transmissive regime is reached we numerically investigate the linear and quadratic OM couplings allowing the membranes in the defect region to move while fixing the side membranes. The shape of the coupling throughout the defect region exhibits indeed the sinusoidal shape characteristic of the transmissive regime as outlined in the previous section. However, the large enhancement with respect to an equidistant case (as tested on the previous OM platform) is not achieved here; we instead find very high values for the couplings which are simply consistent with the extreme localization of light within the small space of dimension $D$. Let us exemplify this by considering a crystal consisting of 47 membranes of polarizability $\zeta=\zeta_m=-0.5$ (with the central 7 making up the defect area with $\alpha=10^{-3}$) and $d=500$ nm. We first tune the resonance by optimizing over $d_m$ which we fix to $247$ nm (according to the procedure described in Fig.~\ref{fig:fig7}). We then achieve an overall coupling of $1.98$ MHz, close to the one expected from the localization of the light mode within a linear dimension of $3.5$ $\mu$m [$(c k/D)x_0 = 2.9$ MHz]. For the quadratic coupling, the analytical estimate for a cavity length of $3.5$ $\mu$m is $0.025$ Hz (the numerical value lies at $0.023$ Hz).

The immediate explanation for the poor performance of transmissive method applied to the OM crystal can be found in the effect of cavity linewidth narrowing as explained in Refs.~\cite{xue12,xue13}. More specifically, the enhancement owed to the access of transparency points is valid only in the regime where the effective cavity optical length $D+(2/\pi^2) d\zeta^2 N^3$ is close to the physical length $D$. When $(2/\pi^2)\zeta^2 N^3$ is non negligible, the predicted scaling of the linear coupling [see Eq.~\eqref{eq:g1simple}] as $\zeta^2 N^{3/2}$ is not valid anymore. Notice that for the chosen example, the condition reads approximately: $2(\zeta N/\pi)^2 <1$; for $\zeta=-0.5$, an ensemble with $N=5$ already breaks the approximation.

\subsection{Discussion of mode structure in the presence of defects }
\label{sec:outlook}
Before concluding, in the following we shortly discuss the qualitative structure of the light modes inside the cavity. This should contribute an intuitive understanding of light localization in the presence of the spatial defects described above. In turn, the latter may be used to design defect configurations that maximize OM couplings, as described above.

\noindent In order to better understand the localization of the light modes inside the cavity array, we note that the Helmoltz equation can be recast in the form of a Schr\"odinger equation for a particle in a 1D periodic potential (see also App.~\ref{sec:appB}). In particular, since we have assumed in Eq.~\eqref{eq:eq2} that the membranes are infinitely thin, our model in the absence of the quadratic defect ($\alpha=0$) is found to correspond to the following Kronig-Penney model (see also App.~\ref{sec:kro})
\begin{equation}\label{eq:eqSchr}
\left[H_0+V^{(\text{eff})}(\bar{x})\right]E(\bar{x},\bar{t})=i\frac{dE(\bar{x},\bar{t})}{d\bar t},
\end{equation}
with $V^{(\text{eff})}(\bar{x})=0$ for $\alpha=0$. Here, the Hamiltonian term $H_0$ reads
\begin{equation}
H_0=-\partial_{\bar{x}^2} - \beta\sum_{i\in\mathbb{Z}}\delta(\bar{x}-\bar{x}_i),
\end{equation}
where $\beta=2(\omega/c)^2(\zeta d/k)$ and the term $\delta(\bar{x}-\bar{x}_i)$ describes the position of the $i^{\rm{th}}$ membrane. The quantities $\bar{x}=x/d$ and $\bar{t}=t J/h$ are the dimensionless spatial and temporal coordinates, respectively, with $J$ the characteristic kinetic energy  of the system and $h$ the Planck's constant. For $\alpha=0$, Eq.~\eqref{eq:eqSchr}  is exactly solvable \cite{ber12} and captures the formation of the band structure, similar to that observed in previous sections.

In the following we assume a well formed band structure and focus on the dynamics in the lowest band, by expanding the electric field $E(\bar{x},\bar{t})$ in terms of first-band Wannier functions only~\cite{rey05}. Analytical expressions for the Wannier functions of the Kronig-Penney are known \cite{ped91} (see also App.~\ref{sec:kro} for a short review), and are similar to exponentially localized position eigenstates. The kinetic energy $J$ above thus corresponds to a fourth of the bandwidth, which can be computed directly from Eq.~\eqref{eq:eqSchr} using the lowest-band Wannier functions (or simply read off directly from the band structure).

For finite $\alpha$, the quadratic defect can be introduced heuristically by adding a term $V^{(\text{eff})}(\bar{x})=\Omega \bar{x}^2$ in Eq.~\eqref{eq:eqSchr}, with $\Omega=\alpha \beta$ the strength of the potential. In this case, the trapping potential is an inverted parabola (that is, it opens downward). Similarly to the problem of a particle in a parabolic potential, solutions of Eq.~\eqref{eq:eqSchr} are immediately recognized as similar to Hermite-type polynomials.
We can gain further insight into the structure of the solutions of  Eq.~\eqref{eq:eqSchr} by noticing that the quantum mechanical problem of a particle in a parabolic potential (with positive curvature) in the discrete tight-binding limit considered here is also exactly solvable~\cite{aun03} in terms of Mathieu functions.
In analogy to the derivation of Ref.~\cite{rey05}, here we obtain that for $4J/\Omega\gg1$, two classes of energy eigenmodes are present: low-energy modes (with energy $E\lesssim 4J$) are close to position eigenstates, i.e., localized on either side of the inverted harmonic potential induced by the defect. We note that this localization is a consequence of the combination of the external confinement due to the parabolic potential and of an inner confinement due to Bragg scattering caused by the periodic membrane potential.  High-energy modes (with energy $E \gtrsim 4J$), instead, are well approximated by harmonic oscillator eigenstates for the inverted parabolic potential. These modes are thus localized around the center of the parabolic defect, as expected. We find this description of the mode structure to be in qualitative agreement with the numerical and analytical findings of previous sections.

While here we have focused only on spatial defects with a quadratic shape, we note that other defect configurations may also lead to large OM couplings. For example, a large localization of the light modes can be obtained by a simple uniform shift of the position of a few membranes in the middle of the array, generating localized modes in the band gaps, similar to what is routinely done in, e.g., photonic band gap systems. In principle, superlattices of these localized defects may be engineered by periodic spatial repetitions of individual defects, allowing in principle for light-induced interactions between membrane modes at different defect positions.

\section{Concluding remarks}
\label{sec:conclusions}

Transmissive OM allows for achieving linear coupling strengths far above those permitted by reflective
OM. This was theoretically proposed in Ref.~\cite{xue12} for an array of equidistant membranes in a high-finesse cavity. In view of the recent experimental progress on OM crystals with quadratic defects~\cite{eic09,cha09}, we have generalized the analysis of Refs.~\cite{xue12,xue13} to include engineered quadratic spatial defects that further enhance both linear and quadratic couplings in the transmissive regime. We have also treated a 1D model for an OM crystal, where we have analytically showed how to reach the transmissive regime and concluded that, owing to the typical small size of the crystal, further enhancement by this technique is not possible. In the last part of the paper we draw an analogy between membrane arrays with quadratic defect and ultracold atoms in an optical lattice plus parabolic potential.

\section{Acknowledgements}
We are grateful to A. Xuereb and A. Dantan for useful discussions. We acknowledge support from the Austrian Science Fund (FWF):\ P24968-N27 (C.~G.),  ERC-St Grant ColdSIM (No. 307688), EOARD (E.~T. and G.~P.), the Universit\'{e} de Strasbourg through Labex NIE and IdEX, the JQI, the NSF PFC at the JQI, Initial Training Network COHERENCE, computing time at the HPC-UdS.

\appendix
\section{First-order expansion}
\label{sec:appA}
Inserting Eq.~(\ref{eq:Fj}) without
$O(\alpha^2)$-terms, Eq.~(\ref{eq:ar}) becomes:

\begin{equation}
\label{eq:exp1}
M_{ar}=M\cdot\left[\prod_{j=1}^{N-1}\left[(\mathbb{I}-ik\alpha
\mathrm{d}_{j,j+1}\sigma_3)\cdot F\cdot M\right]\right].
\end{equation}

In the next subsections we consider $N=3$ and $N=4$ and extend the method for a larger
number of membranes.

\subsection{Three-membrane array}

\noindent For $N=3$ Eq.~(\ref{eq:exp1}) is
\begin{equation}
\label{eq:matrixp1} M_{ar}=M\cdot(\mathbb{I}-ik\alpha
\mathrm{d}_{1,2}\sigma_3) \cdot F\cdot M\cdot (\mathbb{I}-ik\alpha
\mathrm{d}_{2,3}\sigma_3) \cdot F \cdot M
\end{equation}
where $\mathrm{d}_{1,2}=\mathrm{d}_{2,3}$. By expanding the product
up to order $O(\alpha^2)$ one gets
\begin{equation}
\begin{split}
\label{eq:sqb}
M_{ar}=M_3-ik\alpha \mathrm{d}_{1,2}[&M\cdot \tilde{F}\cdot M\cdot F\cdot M+\\
&M\cdot F\cdot M \cdot\tilde{F}\cdot M],
\end{split}
\end{equation}
with $\tilde{F}=F\cdot\sigma_3=\sigma_3\cdot F$. Matrix $M_3$ is in
the absence of defect $M_3=M\cdot F\cdot M\cdot F\cdot M$.
In general $M_N$ can be worked out by following
Refs.~\cite{xue12,xue13}, and is given by Eq.~(\ref{eq:paraform}). To
evaluate the terms within square brackets in Eq.~(\ref{eq:sqb}) we first
perform the substitution $\tilde{F}=F_{\frac{1}{2}}\cdot\tilde{F}_{\frac{1}{2}}$,
where $F_{\frac{1}{2}}$ is the matrix describing propagation of
a monochromatic beam of wave number $k$ over a distance $d/2$ through
free space:
\begin{equation}
F_{\frac{1}{2}}=
\twotwomat{e^{ik\frac{d}{2}}}{0}{0}{e^{-ik\frac{d}{2}}},
\end{equation}
and
$\tilde{F}_{\frac{1}{2}}=F_{\frac{1}{2}}\cdot\sigma_3=\sigma_3\cdot
F_{\frac{1}{2}}$. Then, we rewrite the free-space matrix on the left
side on $\tilde{F}$ as $F=F_{\frac{1}{2}}\cdot F_{\frac{1}{2}}$,
whereas that on the right as $F=\tilde{F}_{\frac{1}{2}}\cdot\tilde{F}_{\frac{1}{2}}$
and eventually we multiply the brackets by the identity matrices $I=F_{\frac{1}{2}}^{-1}\cdot F_{\frac{1}{2}}$
on the left side, and $I=\tilde{F}_{\frac{1}{2}}\cdot\tilde{F}_{\frac{1}{2}}^{-1}$ on right side. We obtain
\begin{equation}
\label{eq:even} M_{ar}=M_3-ik\alpha
\mathrm{d}_{1,2}F_{\frac{1}{2}}^{-1}\left[A\cdot {A^t}^2+A^2\cdot
A^t\right]\tilde{F}_{\frac{1}{2}}^{-1}
\end{equation}
where
\begin{equation}
\label{eq:M1} A=F_{\frac{1}{2}}\cdot M\cdot F_{\frac{1}{2}}.
\end{equation}
The matrices $A$ (and $A^t$) are unimodular.

\subsection{Four-membrane array}
\noindent For $N=4$
\begin{equation}
\label{eq:matrixp1}
\begin{split}
M_{ar}=&M\cdot(\mathbb{I}-ik\alpha \mathrm{d}_{1,2}\sigma_3) \cdot F\cdot M\cdot (\mathbb{I}-ik\alpha \mathrm{d}_{2,3}\sigma_3) \cdot F \cdot \\
&M \cdot (\mathbb{I}-ik\alpha \mathrm{d}_{3,4}\sigma_3) \cdot F
\cdot M,
\end{split}
\end{equation}
where $\mathrm{d}_{1,2}=\mathrm{d}_{3,4}$. By expanding up to order
$O(\alpha^2)$ one obtains:
\begin{equation}
\begin{split}
M_{ar}=&\;M_4-ik\alpha \mathrm{d}_{1,2} [M\cdot\tilde{F}\cdot M\cdot F \cdot M\cdot F\cdot M+\\
&M\cdot F \cdot M\cdot F\cdot M \cdot\tilde{F}\cdot M]-\\
&ik\alpha \mathrm{d}_{2,3} [M\cdot F\cdot M\cdot\tilde{F}\cdot
M\cdot F\cdot M]
\end{split}
\end{equation}
where $M_4$ is  without defect. With a bit of more manipulations
\begin{equation}
\begin{split}
M_{ar}=M_4-ik\alpha F_{\frac{1}{2}}^{-1}\cdot [&\mathrm{d}_{1,2}\left(A\cdot{A^t}^3+A^3\cdot A^t\right)+\\
&\mathrm{d}_{2,3}\cdot A^2\cdot{A^t}^2]\cdot\tilde{M}^{-1}_p(d/2).
\end{split}
\end{equation}
If we symmetrize $A^2\cdot{A^t}^2$,
$(A^2\cdot{A^t}^2+A^2\cdot{A^t}^2)/2$, then
\begin{equation}
\begin{split}
\label{eq:odd}
M_{ar}=&M_4-ik\alpha\sum_{j=1}^{2}\mathrm{d}_{j,j+1}\left(1-\frac{\delta_{j,2}}{2}\right)\times\\
&F_{\frac{1}{2}}^{-1}
\cdot[A^j\cdot {A^t}^{4-j}+{A}^{4-j}\cdot {A^t}^{j}]\cdot\tilde{F}_{\frac{1}{2}}^{-1}.\\
\end{split}
\end{equation}

\subsection{General case}
\noindent For a generic $N$, the straightforward extension of Eqs.~(\ref{eq:even}) and (\ref{eq:odd}) yields
\begin{equation}
\label{eq:general}
\begin{split}
M_{ar}=&M_N-ik\alpha\sum_{j=1}^{[N/2]}\mathrm{d}_{j,j+1}\left(1-\frac{\delta_{j,\frac{N}{2}}}{2}\right)\times\\
&F_{\frac{1}{2}}^{-1} \cdot[A^j\cdot {A^t}^{N-j}+ {A}^{N-j}\cdot
{A^t}^{j}]\cdot\tilde{F}_{\frac{1}{2}}^{-1}.
\end{split}
\end{equation}

Since both $A$ and $A^t$ are unimodular matrices, we can use
Chebyshev's identity to express their powers \cite{yeh05}. As a
consequence, matrix $C=A^j\cdot{A^t}^{N-j}+A^{N-j}\cdot{A^t}^j$ has the
following elements:
\begin{equation}
\label{eq:C}
\begin{split}
[C]_{11}&= [C]^*_{22}=2\left[\mathbb{C}_{j-1}\mathbb{C}_{N-j-1}-{\zeta}^2 U_{j-1}(a)U_{N-j-1}(a)\right],\\
[C]_{12}&=[C]_{21}=4\zeta b U_{j-1}(a)U_{N-j-1}(a),
\end{split}
\end{equation}
where
\begin{empheq}[left=\empheqlbrace]{align}
\label{eq:a}
&a(x)=\cos(x)-\zeta\sin(x)\\
\label{eq:b}
&b(x)=\sin(x)+\zeta\cos(x)\\
&\mathbb{C}_i(x)=\left(U_{i-1}(a(x))-e^{ix}(1+i\zeta)U_{i}(a(x))\right).
\end{empheq}
In Eqs.~(\ref{eq:C}) we wrote $a$, $b$, and $\mathbb{C}_i$,
instead of $a(kd)$, $b(kd)$, and $\mathbb{C}_i(kd)$. Function $a(x)$
is the same as in Ref.~\cite{xue12,xue13}, whereas $b(x)$ is nothing
but $a(x-\pi/2)$. $U_j$ is the Chebyshev's polynomial of the second
kind and degree $j$. Product
$F^{-1}_{\frac{1}{2}}C\tilde{F}^{-1}_{\frac{1}{2}}$, in Eq.~(\ref{eq:general}) is
\begin{equation}
F^{-1}_{\frac{1}{2}}C\tilde{F}^{-1}_{\frac{1}{2}}= \begin{bmatrix}
[C]_{11}e^{-ikd} & -[C]_{12} \\ [C]_{12} & -[C]^*_{11}e^{ikd}
\end{bmatrix}.
\end{equation}
By inserting this expression in Eq.~(\ref{eq:general}) we obtain
$\alpha$-expansion given by Eq.~(\ref{eq:Mtot}).

\section{The Schr\"odinger-type equation}
\label{sec:appB}
The Helmoltz equation of Eq.~\eqref{eq:waveq} can be mapped onto the Schr\"odinger equation for an electron in the 1D periodic potential corresponding to 1D tight-binding model.
\begin{equation}
\label{eq:schrodi}
\left[-\frac{\hbar^2}{2m}\partial_x^2+V(x)\right]E(x)=\mathcal{E} E(x)
\end{equation}
with
\begin{empheq}[left=\empheqlbrace]{align}
\label{eq:energia}
&\mathcal{E}=\frac{\hbar^2}{2m}\left(\frac{\omega}{c}\right)^2,\\
\label{eq:potenziale}
&V(x)=-\frac{\hbar^2}{2m}\left(\frac{\omega}{c}\right)^2\delta\epsilon_r(x).
\end{empheq}
Here $m=\hbar^2/(2J d^2)$ with $4J$ the width of the lowest energy band. If $x\not=0$ the potential $V(x)$ vanishes and
Eq.~(\ref{eq:schrodi}) reduces to the wave equation for propagation
through vacuum:
\begin{equation}
\left[\partial^2_x+\left(\frac{\omega}{c}\right)^2\right]E(x)=0.
\end{equation}
In that case if we assume $E(x)=A e^{ikx}$ with $A$ a complex
amplitude, we obtain the usual linear dispersion relation for an
electromagnetic wave propagating through vacuum, $\omega=k c$.

For an infinite array of beam splitters, $V(x)$ is just the direct
extension of (\ref{eq:potenziale}):
\begin{equation}
\label{eq:uneffected}
V(x)=-\left(\frac{\omega}{c}\right)^2\left(\frac{\zeta}{k}\right)\sum_{i\in\mathbb{Z}}\delta(x-x_i).
\end{equation}
The Dirac deltas are centered at the positions of the elements and the
quadratic spacing gradient of Eq.~(\ref{eq:quadpos}) only acts on the
$x_i$'s. We can express the periodic Dirac comb with period $d$ by using the
Fej\'er kernel:
\begin{equation}
\label{eq:fej}
\delta_{\text{comb},d}(x)=\lim_{N\to\infty}\frac{1}{dN}\frac{\sin{\left(\frac{N\pi}{d}x\right)}^2}{\sin{\left(\frac{\pi}{d}x\right)}^2}.
\end{equation}
For a quadratic defect [see Eq.~(\ref{eq:quadpos})] over a length $D$ of the infinite array, the
corresponding Dirac comb is
\begin{equation}
\begin{split}
\label{eq:fej1}
\delta_{\text{comb}}(x)=&\lim_{N\to\infty}\frac{1}{dN}\times\\
&\frac{\sin{\left(\frac{N\pi}{d}\left(x-\frac{d}{2}+\frac{\alpha}{d}\left[\frac{D^2}{4}-x^2\right]\text{Sgn}(x)\right)\right)}^2}{\sin{\left(\frac{\pi}{d}\left(x-\frac{d}{2}+\frac{\alpha}{d}\left[\frac{D^2}{4}-x^2\right]\text{Sgn}(x)\right)\right)}^2},
\end{split}
\end{equation}
which holds true as long as $\alpha<2d^2/{(D^2-d^2)}$. A finite size $L$ of the array is straightforwardly taken into
account by means of a proper  Heaviside-step-functions combination.
For a total length $L$ of the crystal and a defect region extended
over a region of size $D<L$, the Schr\"odinger equation is
Eq.~(\ref{eq:schrodi}) with a potential term $V(x)$ given by
Eq.~(\ref{eq:fej1}) multiplied by $\theta(x+L/2)-\theta(x-L/2)$.


\section{Discussions}
\label{sec:appendixC}
\label{sec:sim} The quadratic defect forms a local effective
potential for optical modes, with the spatial dependence of the
effective potential closely following the spatial properties of the
defect itself~\cite{eic09,cha09}. Optical modes of the
infinitely-periodic structure are confined by a quasi-harmonic
potential which is concave downward. This effective potential
localizes a ``ladder'' of modes with Hermite-Gauss envelopes,
analogous to the modes of the $1$D harmonic potential; higher-energy modes will be those more localized in the
middle of the quasi-periodic region. Had we introduced an inverted
defect, keeping the total length constant (that
is fixing the positions for membranes $1$ and $N$) and pushing
the membranes towards the outside, the effective potential would have
changed curvature and higher-energy modes would have then been those less localized. This can be seen in Fig.~\ref{fig:fig8}
where the $N-1$ resonances belonging to the first transmissive
band are shown for eight membranes with polarizability
$\zeta=-4$ and a defect $\alpha=3\cdot10^{-3}$. The highest energy mode, with seven (i.e. $N-1$)
nodes, is the most localized (red dashed
curve) because, the defect immediately pushes it inside the
transmissive gap, see Secs.~\ref{subsec:refb}.

One can think of photons on the OM platform as particles in connected boxes. The length of the box defines the spectrum, and shortening (increasing) the distance between two
 membranes modifies the available energies of the respective boxes.

\begin{figure}[t]
\includegraphics[width=0.99\columnwidth]{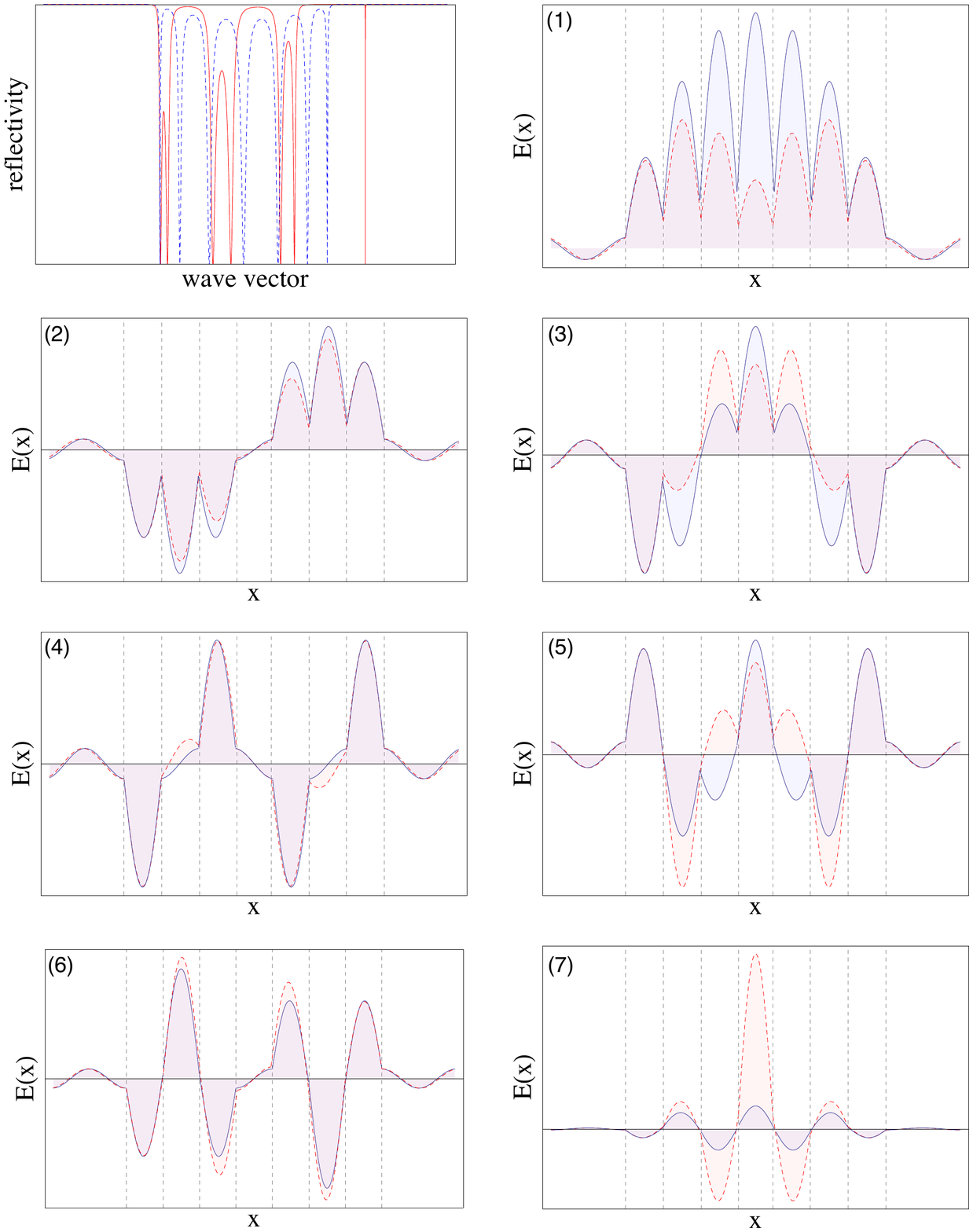}
\caption{\textit{Light localization}. The optical response is plotted in the uppermost right plot as a function of incoming wave vector for $\zeta=-4$ and $N=8$ membranes for both equidistant (blue line) and defect case with $\alpha=3\cdot 10^{-3}$. For each of the $N-1$ transmissive points, we plot the corresponding electric field amplitude profile (the plot label corresponds to the transmission order) along the array with its defect induced localization.} \label{fig:fig8}
\end{figure}

\section{The Kronig-Penney model}
\label{sec:kro}

\noindent The Schr\"odinger equation for an infinite photonic
crystal without parabolic potential reduces to
\begin{equation}
\label{eq:basicsc}
\left[-\frac{\hbar^2}{2m}\partial_x^2+V(x)\right]E(x,t)=i\hbar\frac{dE(x,t)}{dt}.
\end{equation}
The potential $V(x)$ is given by Eq.~(\ref{eq:uneffected}). In
Eq.~(\ref{eq:basicsc}) we simply have the Hamiltonian for the
1D Kronig-Penney model, with inter-membrane spacing $d$
and $\delta$-walls at
\begin{equation}
x_j=\left(j-\frac{1}{2}\right)d\;\,\mbox{with}\;\;j\in\mathbb{Z}.
\end{equation}
Equations~\eqref{eq:basicsc}, \eqref{eq:uneffected} explain why the polarizability $\zeta$ has to be negative when we
are dealing with mirrors: if it was positive each mirror would
behave as an infinite well, corresponding to an attractive
$\delta$-potential. The Schr\"odinger equation (\ref{eq:basicsc})
would then allow for bound states in the lowest (zero-th) band among
its solutions, but a mirror can not trap photons.

The solutions of Eq.~(\ref{eq:basicsc}) are Bloch functions~\cite{ped91}
\begin{equation}
\psi_{n,q}(x)=e^{iqx}u_{n,q}(x).
\end{equation}
In the interval $jd<x\leq (j+1)d$ they change according to:
\begin{equation}
\psi_{n,q}^{(j)}(x)=e^{iqjd}\psi^{(0)}_{n,q}(x-jd).
\end{equation}
Between $\delta$-walls the wave-function satisfies the free-space
Schr\"odinger equation. Since we are dealing with $\delta$-walls only
positive energies are allowed:
\begin{equation}
\label{eq:eenne1} \mathcal{E}=\frac{\hbar^2 k^2}{2md^2}\geq0.
\end{equation}
By comparing Eqs.~(\ref{eq:energia}) and (\ref{eq:eenne1}) it is
apparent that $k=\omega c$. Unlike $q$, which is the quasi-momentum
of the Bloch wave, $k$ is the real wave vector of the optical modes
confined within the membranes and thus follows from the band
structure. Momentum $k$ and quasi-momentum $q$ are related via the
dispersion relation of Eq.~(\ref{eq:dispersionrelation}).

\noindent According to \cite{ped91}, in the interval $-d/2<x\leq d/2$
\begin{equation}
\begin{split}
\psi_{n,q}^{(0)}(x)=A[&\cos{(qd/2)}\sin{(kd/2)}\cos{kx}+\\
&i\sin{(qd/2)}\cos{(kd/2)}\sin{kx}],
\end{split}
\end{equation}
where the modulus of $A$ is determined by the renormalization condition:
\begin{equation}
\int_{\tiny{\mbox{cell}}}|\psi_{n,q}(x)|^2dx=\int_{-d/2}^{d/2}|\psi_{n,q}^{(0)}(x)|^2dx=1.
\end{equation}
The result is
\begin{equation}
\frac{4}{d}|A|^{-2}=\sin{(kd)}-\frac{\zeta}{kd}\sin{(kd)}(\sin{(kd)-kd\cos{(kd)}}).
\end{equation}
The Wannier functions are then given by
\begin{equation}
w_{n,j}(x)=\frac{d}{2\pi}\int_{-\pi/d}^{\pi/d}e^{iqjd}\psi^{(0)}_{n,q}(x-jd)dx.
\end{equation}
The theory of Wannier functions is complicated by the presence of a
``gauge freedom" that exists in the definition of the Bloch waves
\cite{mar12}. Different choices of smooth gauge correspond to
differents sets of Wannier functions having in general different
shapes and spreads.

\section{Band theory}

\begin{figure}[t]
\includegraphics[width=0.99\columnwidth]{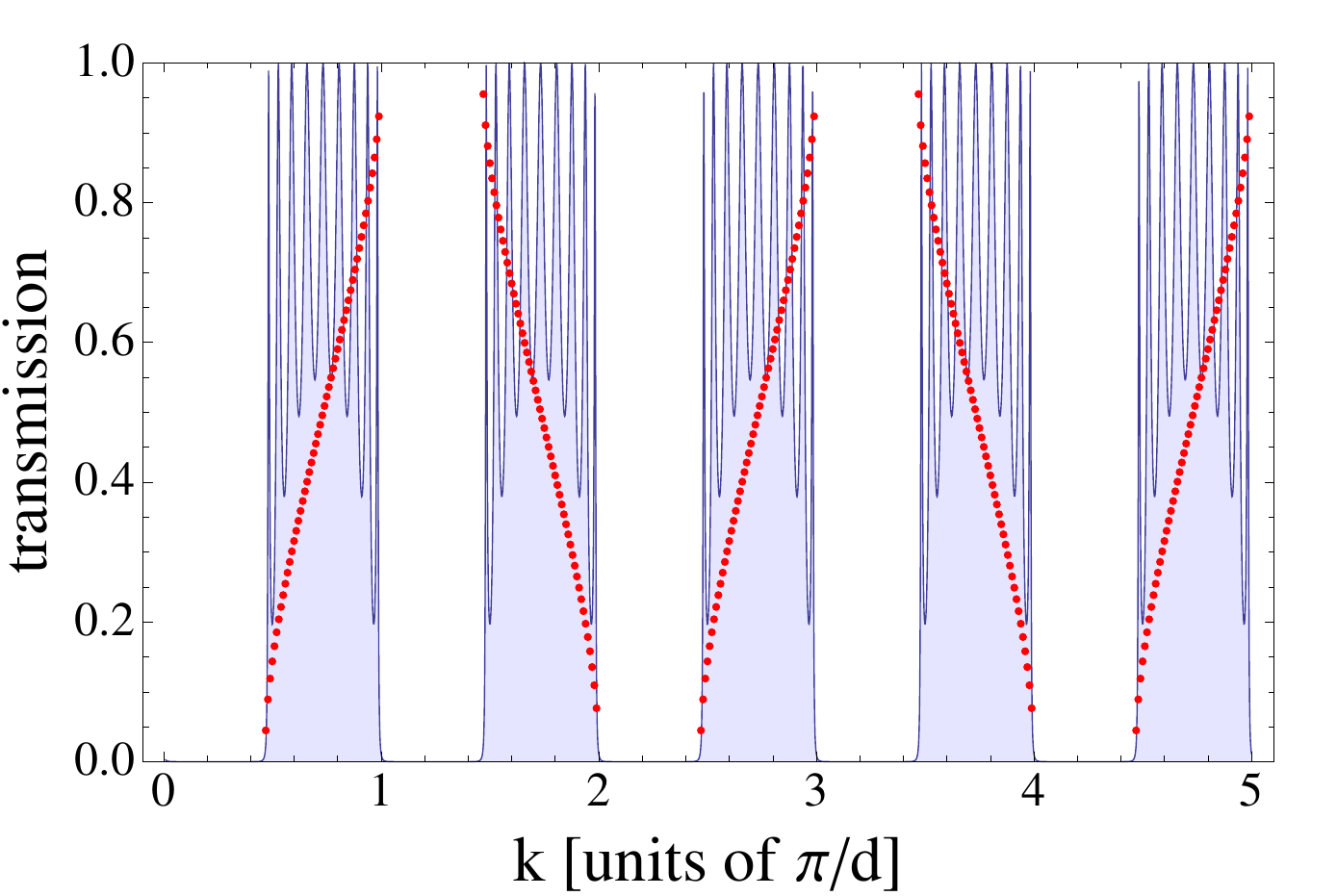}
\caption{Overlap between the transmission plot (blue) for a finite
size array of membranes and the band structure (red) of an infinite
1D photonic periodic crystal. Ten membranes are considered. Both
plots have a polarizability $\zeta=-0.9$.} \label{fig:fig9}
\end{figure}

\label{sec:appE} The product $M\cdot F$ of a matrix of
Eq.~\eqref{eq:M} with $\zeta_i=\zeta$ and a free-space matrix of
Eq.~(\ref{eq:F}) with $x_{j,j+1}=d$ is the transfer matrix for a
modular element of the infinite array. It is a unimodular matrix and
has a real trace, see Refs.~\cite{mar08,joa08}. Its two eigenvalues
are related by
\begin{equation}
\lambda_2=\frac{1}{\lambda_1}.
\end{equation}
If $|\lambda_1|=1$ it can be written as
\begin{equation}
\lambda_1=e^{iqd},
\end{equation}
with $q$ being real. As $\lambda_2=e^{-iqd}$, $\text{Tr}\,[M\cdot
F]=2a= \lambda_1+1/\lambda_1=2\cos{qd}$, where $a$ is function of
$kd$ and is defined in Eq.~(\ref{eq:a}). Note that
\begin{equation}
\label{eq:traccia} |\mbox{Tr}\, [M\cdot F]|\leq 2.
\end{equation}
If $|\lambda_1|\not=1$, instead, $|\mbox{Tr}\; [M\cdot F]|
=|2a|=|\lambda_1+\lambda_1^{-1}|=2\cosh{\kappa d}>2$, and the
amplitude of the transmitted wave decreases exponentially with
increasing width of the membrane. Eq.~(\ref{eq:traccia}) represents
a sufficient condition for the existence of propagating solutions.
Bloch wave vector $q$ and $k$ are related by $\cos{qd}=a$:
\begin{equation}
\label{eq:dispersionrelation}
q=\frac{1}{d}\arccos{a}=\frac{1}{d}\arccos{(\cos{kd}-\zeta\sin{kd})}.
\end{equation}

For negative $\zeta$ (repulsive potential) the top of the $n$-th
band corresponds to $k=(n+1)\pi/d$ ($n=0,1,2,\dots$) and the lowest
band starts from a strictly positive $k$. For a finite size every
band turns into a band containing $N-1$ resonances. The band
structure gives us some hints about the transmission plot: the first
resonance corresponds to a collective optical mode without nodes,
the second one has one node, and so on. The width of the band-gap
behaves as $\exp(\zeta)$, and width of the band as $\exp(-\zeta)$.

\end{document}